\documentclass[10pt, a4paper]{article}

\usepackage[utf8]{inputenc}    
\usepackage[T1]{fontenc}       
\usepackage{lmodern}            
\usepackage[title]{appendix}
\usepackage{booktabs} 
\usepackage{graphicx}
\usepackage{geometry}
\usepackage{authblk}    
\usepackage{cite}       
\usepackage{amsmath}    
\usepackage{amssymb}    
\usepackage{amsfonts}   
\usepackage{mathtools}
\usepackage{mathrsfs}
\usepackage{url}        
\usepackage[colorlinks=true, linkcolor=blue, citecolor=blue, urlcolor=blue]{hyperref} 

\title{Suppressing secondary shock waves in jam-absorption driving via string-stable support vehicles}

\author[1]{Atsushi Suzuki}
\author[2]{Akihiro Tokumitsu}
\author[2, *]{Ryosuke Nishi}

\affil[1]{Department of Engineering, Graduate School of Sustainability Science,
Tottori University, 4-101 Koyama-cho Minami, Tottori, Tottori 680-8552, Japan}

\affil[2]{Department of Mechanical and Physical Engineering, Faculty of Engineering,
Tottori University, 4-101 Koyama-cho Minami, Tottori, Tottori 680-8552, Japan}

\affil[*]{Corresponding author: \href{mailto:nishi@tottori-u.ac.jp}{nishi@tottori-u.ac.jp}}

\date{\today}

\begin{document}

\maketitle

\begin{abstract}
As a freeway-driving strategy, jam-absorption driving (JAD) clears a traffic shock wave (stop-and-go wave) by slowing down a single vehicle, called the absorbing vehicle. However, JAD may destabilize the traffic flow upstream of this vehicle, generating secondary shock waves. This study proposes a method to suppress secondary shock waves by controlling the behavior of connected and automated vehicles (CAVs) upstream of the absorbing vehicle, called support vehicles (SVs). A string-stability-based control method is applied in which SVs dynamically extend their time gaps to provide support driving (SD) for JAD. Numerical simulations revealed that SD damped perturbations caused by the absorbing vehicle and prevented secondary shock waves, consistent with the head-to-tail string stability criterion. Combining JAD and SD reduced fuel consumption and collision risk compared with the JAD-only method, but increased travel time. Reverting the extended time gap to its initial value reduced travel time while maintaining low collision risk compared with the non-reverting method, albeit with increased fuel consumption. Thus, combining JAD and SD effectively eliminates the target shock wave while suppressing secondary shock waves with guaranteed string stability.
\end{abstract}

\vspace{1em}

\noindent\rule{\linewidth}{0.5pt}

\section{\label{sec:introduction}Introduction}
Traffic congestion leads to significant time, fuel, financial, and environmental losses~\cite{Schrank2024}, highlighting the need for effective mitigation. Infrastructural strategies have been developed to resolve freeway congestion using a variable speed limit (VSL)~\cite{Lu2014,Khondaker2015} and/or ramp metering~\cite{Papageorgiou2002}. These strategies control freeway traffic via infrastructural upgrades, such as variable message signs placed at least several hundred meters apart and ramp meters. Seminal strategies include an ramp-metering strategy for preventing mainstream congestion~\cite{Papageorgiou1991}, model predictive control (MPC)-based dynamic VSL strategies for suppressing a traffic shock wave (also known as a moving jam, traffic oscillation, or stop-and-go wave)~\cite{Hegyi2005, Han2017TRC}, a kinematic wave theory-based dynamic VSL strategy named the SPEed ControllIng ALgorIthm using Shockwave Theory (SPECIALIST) for clearing a shock wave~\cite{Hegyi2008, Hegyi2010}, and the mainstream traffic flow control for suppressing capacity drop at bottlenecks~\cite{Carlson2010TranspSci, Carlson2010TRC}.

The development of connected vehicles (CVs), automated vehicles (AVs), and vehicle-to-infrastructure technologies has inspired strategies for directly controlling the maneuver of individual vehicles to alleviate freeway congestion~\cite{Yu2021}. Such vehicle-level strategies do not require expensive infrastructure, such as variable message signs, and are projected to achieve more cost-effective and flexible operation than infrastructure-level strategies. For instance, Kesting et al.~\cite{Kesting2008} developed an adaptive-cruise control (ACC) strategy dynamically tuning the microscopic car-following performance in response to traffic situations, and restricted bottleneck congestion with a $25\%$ penetration rate of the ACC-equipped vehicles. Here, car-following performance refers to how effectively a vehicle adjusts its velocity and spacing in response to the movement of the vehicle directly ahead. Knorr et al.~\cite{Knorr2012} proposed a vehicle-to-vehicle communication-based strategy recommending CVs to enlarge their gaps in response to downstream congestion, and reduced travel time with a $10\%$ or less penetration rate of the CVs. Han et al.~\cite{Han2021} proposed an MPC-based hierarchical connected car-following control (CFC) of connected and automated vehicles (CAVs), and dissipated a shock wave with a $5\%$ penetration rate of the CAVs. Li and Li~\cite{Li2019} proposed a trajectory planning for CAVs, and removed a traffic jam in a fully connected and automated environment. One of the problems with these strategies is that these strategies need some penetration rate of dedicated vehicles (AVs, CVs, or CAVs). When their penetration rate is extremely low, such as when only one dedicated vehicle is present in the system, this vehicle is required to perform a special maneuver to eliminate congestion.

\begin{figure}[t]
\label{fig:problems_and_scope}
\centering
\includegraphics[width=\hsize]{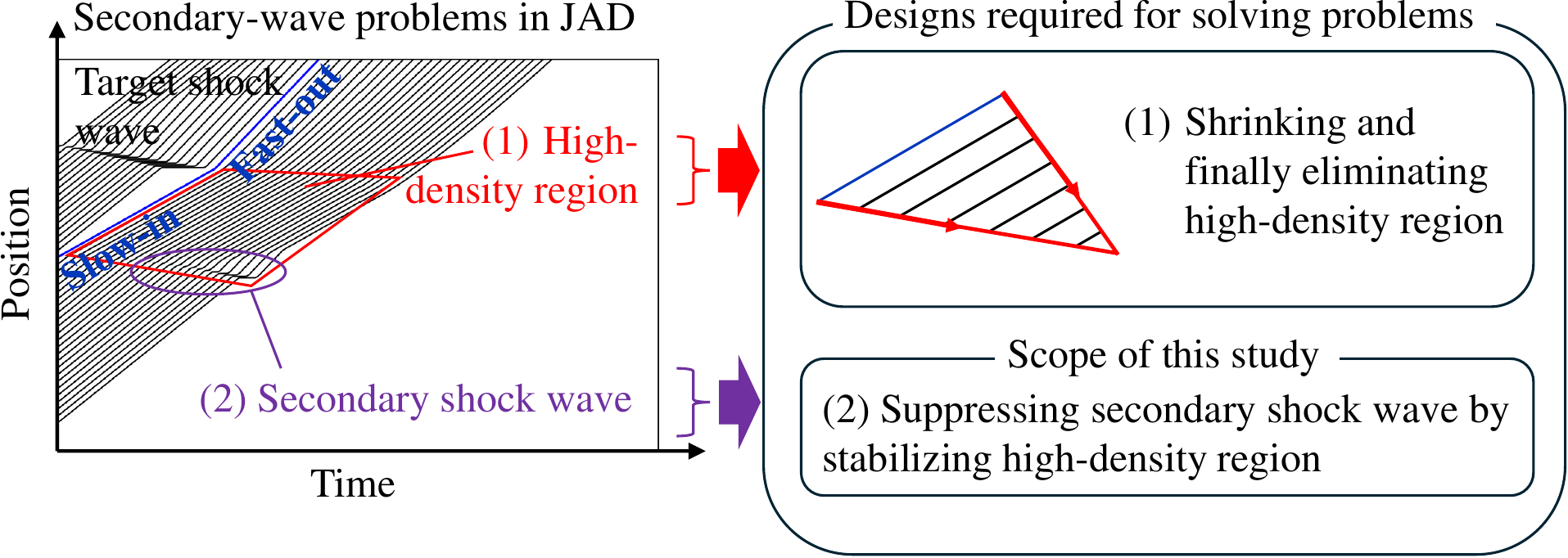}
\caption{(Left) Schematic of secondary waves in jam-absorption driving (JAD), composed of high-density region and secondary shock wave. The blue line indicates a single absorbing vehicle's spatiotemporal trajectory.
(Right) Designs of JAD required for solving the secondary-wave problems. The scope of this study is to suppress the secondary shock wave by stabilizing the high-density region. To shrink and eliminate the high-density region is out of the scope of this study.}
\end{figure}

Researchers have developed driving strategies using special maneuvers of only a single dedicated vehicle~\cite{Beaty1998, Behl2010, Washino2003, Nishi2013, Taniguchi2015, He2017, Han2017TRB, Stern2018, Ghiasi2019, Zheng2020ITJ, Wu2022, Nishi2022, He2025} or a single row of dedicated vehicles~\cite{Ramadan2017, Piacentini2018, Yang2019}, which are not necessarily CAVs, in order to mitigate or clear freeway congestion such as a single shock wave~\cite{Beaty1998, Washino2003, Nishi2013, Taniguchi2015, He2017, Yang2019, He2025} and bottleneck congestion~\cite{Behl2010, Ramadan2017, Han2017TRB, Piacentini2018, Ghiasi2019, Nishi2022} on non-periodic roads, and a single shock wave on ring roads~\cite{Stern2018, Zheng2020ITJ, Wu2022}. Among them, this study focuses on jam-absorption driving (JAD)~\cite{Nishi2013, Taniguchi2015, He2017, He2025}. JAD is performed using a single vehicle (called an absorbing vehicle) to clear a shock wave on a single-lane road. Fig.~\ref{fig:problems_and_scope} illustrates the spatiotemporal trajectory of the absorbing vehicle (blue line). JAD actions are composed of slow-in and fast-out~\cite{Nishi2013}. In the slow-in action, the absorbing vehicle upstream of the target shock wave stops its car-following behavior and runs at a lower velocity (called the absorbing velocity) to avoid being trapped in the shock wave. As no vehicles are added to the shock wave, the shock wave shrinks until it disappears. Subsequently, in the fast-out action, the absorbing vehicle resumes following the vehicle in front of it. The advantages of JAD are as follows: First, JAD action requires only one dedicated vehicle. Second, it does not require enhancing the microscopic car-following performance of vehicles. Conducting JAD should be possible with a patrol vehicle owned by police organizations~\cite{He2025} and/or highway operating companies.

However, fresh congestion generated upstream of the absorbing vehicle, called secondary waves (or secondary traffic jams), poses a major challenge to JAD. Secondary waves can be categorized into three types~\cite{He2025}. The first type is high-density region, where vehicles run at the absorbing velocity (lower than the initial velocity), as shown in Fig.~\ref{fig:problems_and_scope}. Such a high-density region is less stable against perturbations than the initial-density region and is more likely to trigger the second type. The second type is fresh shock wave (secondary shock wave) produced by the unstable traffic flow, as shown in Fig.~\ref{fig:problems_and_scope}. The third type is another fresh shock wave that occurs when the absorbing vehicle and the following vehicles run with a high velocity and small gaps downstream of the first-type high-density region. Since this study does not consider such a small-gap running state, the third type is out of the scope of this study.

The high-density region can be weakened by two methods. The first is to shrink the region until it vanishes~\cite{Nishi2013, He2025}. According to the latest review~\cite{He2025}, this region can be shrunk through fundamental diagrams (flow-density or velocity-gap diagrams) that incorporate capacity drop, similar to the SPECIALIST approach~\cite{Hegyi2008}. The second method involves converting this region into a more stable region. Wang~\cite{Wang2018} numerically damped perturbations propagating through a vehicle platoon composed of six vehicles by gradually extending the time-gap parameter of a single CAV inside the platoon on the basis of a stability margin (i.e., a degree of string stability) of the platoon. Here, string stability refers to the stability determining the growth of a perturbation propagating through a vehicle platoon~\cite{Feng2019}. In JAD scenarios, extending the time gap of dedicated vehicles upstream of the absorbing vehicle should stabilize the vehicle platoon in that region (i.e., make the vehicle platoon in that region string stable) and suppress secondary shock waves.
We define these dedicated vehicles as support vehicles (SVs) and their action as support driving (SD). Although deploying SVs requires dedicated vehicles in addition to the absorbing vehicle, the total number of dedicated vehicles required for JAD and SD should be small.

As explained in detail in the latest review~\cite{He2025}, VSL~\cite{Hegyi2005, Hegyi2008, Wang2012, Wang2014, Wang2016JITS, Han2017TRC, Han2021} and JAD~\cite{Nishi2013, Taniguchi2015, He2017, Nishi2020, Zheng2020AAP, Li2024, Liu2025} strategies have been developed for clearing shock waves.
Utilization of moving bottlenecks (that is, vehicles running more slowly than the surrounding vehicles)~\cite{Yang2019, Cicic2019, Cicic2022TRB} has also been developed for removing or dissipating shock waves. However, to our knowledge, SD with guaranteed string stability has not been applied yet for the suppression of secondary shock waves. Therefore, the aim of this study is constructing SD that realizes string stability for the high-density region, and suppresses secondary shock waves, in a JAD scenario to clear a shock wave. To this aim, it is necessary to investigate the effects of SD on the stability and traffic patterns of the vehicle platoon upstream of the absorbing vehicle. Moreover, it is necessary to investigate the effects of SD in terms of traffic performance indices (such as travel time, fuel consumption, and safety measure), the influence of different SD maneuvers (such as reverting extended time gaps or not~\cite{Wang2018}), and the necessity of the combination of JAD and SD.
\subsection{\label{subsec:contributions}Contributions}
The contributions of this study are as follows:
\begin{enumerate}
\item[(i)] In a JAD scenario (a scenario to clear a shock wave by a single vehicle), we construct a string-stability-based SD to suppress secondary shock waves for the first time, as an application of the dynamic stabilization of vehicle platoons~\cite{Wang2018}.
\item[(ii)] We numerically confirm the impact of SD on the string stability of the vehicle platoon upstream of the absorbing vehicle.
\item[(iii)] We numerically investigate the effects of SD using space-time and velocity-time diagrams and performance indices such as total travel time, total fuel consumption, and total collision risk.
\item[(iv)] We confirm the necessity for SVs to revert their extended time gaps to the initial value~\cite{Wang2018} in the JAD scenario.
\item[(v)] We confirm the need for combining JAD and SD compared with JAD-only and SD-only scenarios.
\end{enumerate}
These contributions will facilitate the realization of flexible and cost-effective restrictions of secondary shock waves in JAD and other vehicle-level strategies.
\subsection{\label{subsec:assumptions}Assumptions}
Real freeway traffic includes vehicles of different types (such as motorcycles, cars, and trucks) and automation and connectivity levels (such as human-driven vehicles (HDVs), AVs, non-CVs, and CVs). Vehicles experience delays~\cite{Wang2018} and errors in their actuators, sensors, and communication devices. HDVs have an intrinsically stochastic nature in their car-following behavior~\cite{Jiang2014, Jiang2015}. Moreover, several freeways have multi-lane roads with curves, slopes, ramps, intersections, and junctions. Furthermore, predicting and estimating shock waves are necessary for their elimination, which requires some time and contains some errors~\cite{Netten2013, Hegyi2013}. Nevertheless, to focus on the fundamental problem of suppressing secondary shock waves, we make the following assumptions:
\begin{enumerate}
\item[(i)] All vehicles have the same length. The vehicles have the same car-following performance except for the leading vehicle, absorbing vehicle, and SVs. The absorbing vehicle and SVs are CAVs. All the other vehicles are human-driven CVs. The actuators or sensors in the vehicles do not exhibit delays or errors. The position and velocity of each vehicle are shared with the other vehicles without delays or errors. None of the vehicles exhibits stochasticity in their car-following behavior.
\item[(ii)] A freeway is a non-periodic and single-lane road without curves, slopes, ramps, intersections, or junctions.
\item[(iii)] The target shock wave is not predicted or estimated. The trajectory of the absorbing vehicle during the JAD actions is predetermined without errors.
\item[(iv)] Suppression of secondary shock waves (that is, stabilization of the high-density region upstream of the absorbing vehicle) is the primary focus. The problem of removing the high-density region~\cite{He2025} is not considered.
\end{enumerate}

The remainder of this paper is organized as follows. Related work is reviewed in Sec.~\ref{sec:related_work}. The proposed method is explained in Sec.~\ref{sec:method}. The numerical results are described in Sec.~\ref{sec:results}. A conclusive discussion is presented in Sec.~\ref{sec:conclusions}.
The nomenclature is listed in Appendix~\ref{sec:nomenclature}.
\section{\label{sec:related_work}Related work}
We briefly review the secondary shock-wave problem in VSL and JAD strategies for eliminating shock waves. In the SPECIALIST field test on the A12 freeway in the Netherlands, the approach generated secondary shock waves in or upstream of the high-density region in $7\%$--$33\%$ of its activations~\cite{Hegyi2010}. The main source of the secondary shock waves observed in the field tests was increased traffic demand~\cite{Han2015}.
Moreover, influence of the traffic demand, speed limit, and VSL duration on the probability of secondary shock waves was numerically investigated under the presence of traffic stochasticity~\cite{Wang2012, Wang2014}.
Numerical simulations of the combination of SPECIALIST and CFC of CAVs also generated secondary shock waves due to the creation of high-density regions by the latter~\cite{Wang2016JITS}.
In JAD, numerical simulations using car-following models revealed that secondary shock waves are more likely to occur as the absorbing velocity decreases~\cite{Taniguchi2015}. Nishi~\cite{Nishi2020} proposed a linear-string-stability-based condition restricting the occurrence of secondary shock waves. Zheng et al.~\cite{Zheng2020AAP} eliminated secondary shock waves using an additional absorbing vehicle. In contrast to these studies, this study stabilizes the high-density region to prevent secondary shock waves from occurring by introducing SD~\cite{Wang2018}.

Researchers have developed driving policies to dissipate a shock wave on a ring road. These included an experimental demonstration of a single AV dissipating a shock wave on a ring road~\cite{Stern2018}, a numerical approach that used reinforcement learning (RL)~\cite{Wu2022}, and an analytical method~\cite{Zheng2020ITJ}. In addition, MPC control policies have been proposed to attenuate shock waves using a platoon of vehicles equipped with ACC or cooperative ACC systems~\cite{Wang2023}. Contrary to the studies on ring roads, this study considers a non-periodic road.

Mitigating freeway bottleneck congestion with longitudinal control of CAVs has garnered considerable research interest. Studies in this direction include the Lagrangian control of moving bottlenecks (e.g., Refs.~\cite{Ramadan2017, Piacentini2018, Cicic2022TITS, Vishnoi2024}), optimal vehicle maneuvers at a sag~\cite{Goni-Ros2016}, and removal of a bottleneck congestion by a single vehicle~\cite{Han2017TRB, Ghiasi2019, Nishi2022}. In contrast to these studies, this study focuses on eliminating a single shock wave.

Researchers have developed driving strategies for weakening single or multiple shock waves using a fraction of dedicated vehicles rather than clearing a single shock wave using a single vehicle. Examples include cooperative CFC with $10\%$ penetration ratio of CAVs~\cite{Wang2016TITS} and hierarchical control of 100 CAVs on a real freeway~\cite{Lee2025}. Notably, an RL controller for AVs alleviated shock waves~\cite{Jang2025}. In contrast to these studies, this study focuses on the removal of a shock wave by a single vehicle. Han et al.~\cite{Han2021} used a hierarchically connected optimal CFC with a CAV penetration rate of $5\%$ to eliminate a shock wave without analyzing the string stability upstream of the CAVs. In contrast to Han et al.'s work~\cite{Han2021}, this study stabilizes the platoon upstream of the absorbing vehicle on the basis of string stability criterion.
\section{\label{sec:method}Method}
\subsection{\label{subsec:system_ini_conds}System description and initial conditions}
\begin{figure}[t]
\centering
\includegraphics[width=\hsize]{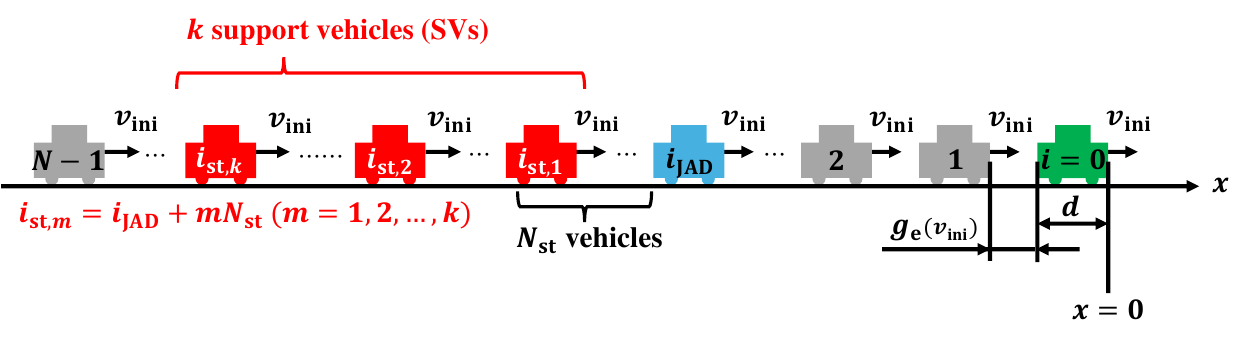}
\caption{Initial conditions of a single-lane system.}
\label{fig:ini_conds}
\end{figure}
This study considers an infinitely long and single-lane freeway as the research object (Fig.~\ref{fig:ini_conds}). At an initial time $t=0\,{\rm s}$, $N$ vehicles are placed on the road with vehicle 0 as the leading vehicle, vehicle $i$ ($i = 1, \ldots, N - 2$) trailing vehicle $i-1$, and vehicle $N-1$ as the last vehicle. No vehicles appear on or disappear from the road during the tests. The forward direction of the vehicles (i.e., the direction of traffic from upstream to downstream) is considered the positive direction of $x$-axis. The front-bumper position and velocity of vehicle $i$ at time $t\,[{\rm s}]$ are denoted as $x_i(t)\,[{\rm m}]$ and $v_i(t)\,[{\rm m}/{\rm s}]$, respectively. Each vehicle has the same length $d\,[{\rm m}]$. The initial position of vehicle 0 is set to $x = 0\,{\rm m}$. Each vehicle has the same initial velocity $v_{\rm ini}\,[{\rm m}/{\rm s}]$. All vehicles, except for the leading vehicle, have the same initial gap (the distance between a vehicle and the one in front of it) $g_{\rm e}(v_{\rm ini})\,[{\rm m}]$, where $g_{\rm e}(v)\,[{\rm m}]$ is the equilibrium gap function at velocity $v\,[{\rm m}/{\rm s}]$. The system has a single absorbing vehicle denoted as vehicle $i_{\rm JAD}$. Every $N_{\rm st}$ vehicle upstream of the absorbing vehicle is designated as an SV.
Parameter $N_{\rm st}$ indicates how many vehicles behind the next SV is from a SV or the absorbing vehicle.
Hence, the number of SVs, $k$, is expressed as
\begin{align}
k= \left\lfloor \dfrac{N - i_{\rm JAD} - 1}{N_{\rm st}} \right\rfloor,
\label{eq:k}
\end{align}
where $\lfloor \cdot \rfloor$ denotes the floor function, such that $\lfloor y \rfloor \coloneqq \max \{ n \in \mathbb{Z} \mid n \le y \}$ for $y \in \mathbb{R}$. The identifier (ID) of the $m$th SV $i_{{\rm st}, m}$ is given by
\begin{align}
i_{{\rm st}, m} = i_{\rm JAD} + m N_{\rm st} \quad\text{for $m = 1, 2, \ldots, k$.}
\label{eq:i_st_m}
\end{align}
We define $A(N_{\rm st})$ as the set of the IDs of SVs:
\begin{align}
A(N_{\rm st}) = \left\{ i_{\rm JAD} + N_{\rm st}, i_{\rm JAD} + 2 N_{\rm st}, \ldots, i_{\rm JAD} + k N_{\rm st} \right\}.
\label{eq:A_N_st}
\end{align}
The parameters are set as $d=5\,{\rm m}$~\cite{Treiber2013}, $N=2000$, and $i_{\rm JAD}=800$. The fixed parameters used in this study are listed in Table~\ref{table:param}.

\begin{table}[t]
\centering
\caption{
Parameter settings.\label{table:param}}
{\begin{tabular}{@{}llllll@{}}
\toprule
Parameter & Value & Parameter & Value & Parameter & Value \\
\midrule
$A\,[{\rm kW} \cdot {\rm s}/{\rm m}]$ & 0.1326 & $N_{\rm st}$ & $100, 200, \ldots, 1000$ & $\alpha_{\rm a}\,[{\rm m}/{\rm s}^2]$ & 1 \\
$a\,[{\rm m}/{\rm s}^2]$ & 1 & $T\,[{\rm s}]$ & 1 & $\alpha_{\rm FR}\,[{\rm g}/{\rm s}]$ & 0.365 \\
$B\,[{\rm kW} \cdot {\rm s}^2/{\rm m}^2]$ & $2.7384 \times 10^{-3}$ & $T_{\rm buf}\,[{\rm s}]$ & 10 & $\alpha_{\rm FR}^{\prime}\,[{\rm g}/{\rm s}]$ & 0.299 \\
$b\,[{\rm m}/{\rm s}^2]$ & 1.5 & $T_{\rm large}\,[{\rm s}]$ & 50 & $\beta$ & 0.08 \\
$C\,[{\rm kW} \cdot {\rm s}^3/{\rm m}^3]$ & $1.0843 \times 10^{-3}$ &  $v_0\,[{\rm m}/{\rm s}]$ & 33.33 & $\beta_{\rm FR}\,[{\rm g}/{\rm m}]$ & $4.10 \times 10^{-3}$ \\
$d\,[{\rm m}]$ & 5 & $v_{\rm a}\,[{\rm m}/{\rm s}]$ & 15.48 & $\gamma_{\rm FR}\,[{\rm g} \cdot {\rm s}/{\rm m}^2]$ & 0 \\
$g_0\,[{\rm m}]$ & 2 & $v_{\rm cr}\,[{\rm m}/{\rm s}]$ & 20.13 & $\Delta t\,[{\rm s}]$ & 0.1 \\
$i_{\rm JAD}$ & 800 & $v_{\rm ini}\,[{\rm m}/{\rm s}]$ & 22 & $\delta$ & 4 \\
$M\,[10^3\,{\rm kg}]$ & 1.325 & $X_{\rm buf}\,[{\rm m}]$ & 100 & $\delta_{\rm FR}\,[{\rm g} \cdot {\rm s}^2/{\rm m}^3]$ & $4.50 \times 10^{-5}$ \\
$N$ & 2000 & $x_{\rm end}\,[{\rm m}]$ & 20000 & $\zeta_{\rm FR}\,[{\rm g} \cdot {\rm s}^2/{\rm m}^2]$ & 0.0943 \\
\bottomrule
\end{tabular}}
\end{table}

To treat the secondary shock wave, we consider a vehicle platoon P composed of all the vehicles upstream of the absorbing vehicle. The leader of this platoon is vehicle $i_{\rm JAD}+1$, the first follower is vehicle $i_{\rm JAD}+2$, while the last follower is vehicle $N-1$. Therefore, the length of platoon P is $N - i_{\rm JAD} - 1$. The absorbing vehicle (vehicle $i_{\rm JAD}$) serves as the exogenous leader of platoon P.
\subsection{\label{subsec:car_following_model}Car-following model}
Each vehicle $i$, except for the leading vehicle or the absorbing vehicle conducting JAD actions, follows vehicle $i-1$ according to the intelligent driver model (IDM)~\cite{Treiber2000}, which is a typical car-following model. The acceleration of vehicle $i$, which is represented as $f(g_i(t), v_i(t), v_{i-1}(t))\,[{\rm m}/{\rm s}^2]$, is determined by its gap $g_i(t)\,[{\rm m}]$, velocity $v_i(t)$, and velocity of the vehicle in front of it, $v_{i-1}(t)$:
\begin{align}
f\left( g_i(t), v_i(t), v_{i-1}(t) \right) = a \left[ 1 - \left\{ \dfrac{ v_i(t) }{ v_0 } \right\}^{\delta} - \left\{ \dfrac{ g^{*}\left( v_i(t), \Delta v_i(t) \right) }{ g_i(t) } \right\}^2 \right],
\label{eq:accel_IDM}
\end{align}
where
\begin{align}
\Delta v_i(t) = v_i(t) - v_{i-1}(t)
\label{eq:Delta_v_i_t}
\end{align}
is the relative velocity (unit: $[{\rm m}/{\rm s}]$) between vehicles $i$ and $i-1$, and
\begin{align}
g^{*}\left( v_i(t), \Delta v_i(t) \right) = g_0 + \max \left( 0, T v_i(t) + \dfrac{v_i(t) \Delta v_i(t)}{2 \sqrt{ab}} \right)
\label{eq:g_star}
\end{align}
is the desired dynamic gap (unit: $[{\rm m}]$)~\cite{Treiber2013}. Parameter $a\,[{\rm m}/{\rm s}^2]$ is the maximum acceleration, $b\,[{\rm m}/{\rm s}^2]$ is the comfortable deceleration, $v_0\,[{\rm m}/{\rm s}]$ is the ideal velocity, $T\,[{\rm s}]$ is the safe time gap, $g_0\,[{\rm m}]$ is the gap in the stopped state, and $\delta$ is the power coefficient. The IDM parameters are set for highway traffic as in Table~11.2 of Ref.~\cite{Treiber2013}, and are listed in Table~\ref{table:param}. The equilibrium gap $g_{\rm e}(v)$ for the IDM is expressed as
\begin{align}
g_{\rm e}(v) = \dfrac{ g_0 + T v }{ \sqrt{1 - \left( \dfrac{v}{v_0} \right)^{\delta} } },
\label{eq:g_e_v_IDM}
\end{align}
which is the solution of $f(g_{\rm e}(v), v, v) = 0$.

In the numerical simulations, vehicle $i$'s position and velocity are updated using the ballistic method~\cite{Treiber2015} at every time interval $\Delta t\,[{\rm s}]$. According to this method, the position and velocity of the vehicle at time $t + \Delta t$ are given by
\begin{align}
x_i\left( t + \Delta t \right) = x_i(t) + v_i(t) \Delta t + \dfrac{f\left( g_i(t), v_i(t), v_{i-1}(t) \right) \left( \Delta t \right)^2}{2}
\label{eq:update_x}
\end{align}
and
\begin{align}
v_i\left( t + \Delta t \right) = v_i(t) + f\left( g_i(t), v_i(t), v_{i-1}(t) \right) \Delta t,
\label{eq:update_v}
\end{align}
respectively.
\subsection{\label{subsec:target_shock_wave}Initial perturbation causing target shock wave}
\begin{figure}[tbp]
\centering
\includegraphics[width=\hsize]{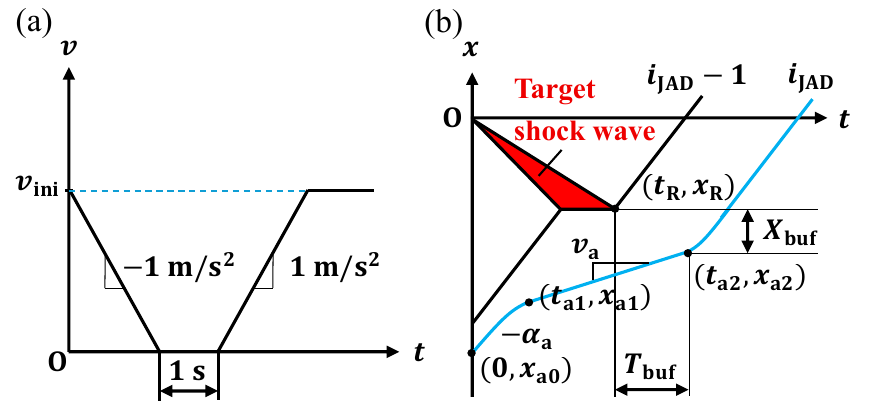}
\caption{(a) Velocity of the leading vehicle (vehicle 0) as a function of time $t$. (b) Schematic of the spatiotemporal trajectory of the absorbing vehicle.}
\label{fig:leading_and_absorbing_vehicles}
\end{figure}
The target shock wave is caused by the initial perturbation of vehicle 0~\cite{Nishi2020}, as illustrated in Fig.~\ref{fig:leading_and_absorbing_vehicles}(a). At $t = 0\,{\rm s}$, vehicle 0 starts decelerating from velocity $v_{\rm ini}$ to $0\,{\rm m}/{\rm s}$ at $-1\,{\rm m}/{\rm s}^2$. After maintaining its velocity at $0\,{\rm m}/{\rm s}$ for $1\,{\rm s}$, it accelerates to $v_{\rm ini}$ at $1\,{\rm m}/{\rm s}^2$. It then maintains its velocity at $v_{\rm ini}$. This perturbation propagates to the following vehicles and evolves into the target shock wave, as shown in Fig.~\ref{fig:xt_vt_3scenarios}(a).
\subsection{\label{subsec:JAD}Jam-absorption driving (JAD)}
Vehicle $i_{\rm JAD}$ performs JAD actions~\cite{Nishi2020}, as illustrated in Fig.~\ref{fig:leading_and_absorbing_vehicles}(b). Hereafter, we express the coordinates in the space-time diagram as $(t\,[{\rm s}], x\,[{\rm m}])$. At the initial space-time point of vehicle $i_{\rm JAD}$, i.e.,
\begin{align}
\left(0, x_{\rm a0} \right) = \left( 0, -i_{\rm JAD} \left( d + g_{\rm e}(v_{\rm ini}) \right) \right),
\label{eq:0_x_a0}
\end{align}
the vehicle starts decelerating from $v_{\rm ini}$ to the absorbing velocity $v_{\rm a}\,[{\rm m}/{\rm s}]$ at $-\alpha_{\rm a}\,[{\rm m}/{\rm s}^2]$, representing the slow-in action. After the velocity reaches $v_{\rm a}$ at
\begin{align}
\left( t_{\rm a1}, x_{\rm a1} \right) = \left( \dfrac{v_{\rm ini} - v_{\rm a}}{\alpha_{\rm a}}, x_{\rm a0} + \dfrac{v_{\rm ini}^2 - v_{\rm a}^2}{2 \alpha_{\rm a}} \right),
\label{eq:t_a1_x_a1}
\end{align}
its velocity is maintained at $v_{\rm a}$ until it reaches $(t_{\rm a2}, x_{\rm a2})$. Because the slow-in action severs the supply of vehicles to the target shock wave, the shock wave shrinks and ultimately disappears. At $(t_{\rm a2}, x_{\rm a2})$, the car-following behavior is initiated according to the IDM, representing the fast-out action. To set the values of $(t_{\rm a2}, x_{\rm a2})$, we define $(t_{\rm R}, x_{\rm R})$ as the space-time point at which vehicle $i_{\rm JAD} - 1$ (i.e., the vehicle in front of vehicle $i_{\rm JAD}$) exits the target shock wave. Point $(t_{\rm R}, x_{\rm R})$ is predetermined by a numerical simulation of the target-shock-wave generation and propagation to the last vehicle without JAD (Fig.~\ref{fig:xt_vt_3scenarios}(a)). In this simulation, $(t_{\rm R}, x_{\rm R})$ is determined by a threshold vehicle-velocity for escaping the shock wave. Since the velocity of vehicles inside the shock wave became zero, the threshold velocity was set to a small and positive value, more precisely, $1\,{\rm m}/{\rm s}$. Using $(t_{\rm R}, x_{\rm R})$, we set $(t_{\rm a2}, x_{\rm a2})$ to
\begin{align}
\left( t_{\rm a2}, x_{\rm a2} \right) = \left( t_{\rm R} + T_{\rm buf}, x_{\rm R} - X_{\rm buf} \right),
\label{eq:t_a2_x_a2}
\end{align}
where $T_{\rm buf}\,[{\rm s}]$ and $X_{\rm buf}\,[{\rm m}]$ denote the time and space buffers, respectively. In the parameter settings listed in Table~\ref{table:param}, the absorbing velocity $v_{\rm a}$ is given by the positive solution of $v_{\rm a}^2 + 2 c_1 v_{\rm a} - c_2 = 0$~\cite{Nishi2020}:
\begin{align}
v_{\rm a} = \dfrac{c_2}{c_1 + \sqrt{c_1^2 + c_2}} = 15.48\,{\rm m}/{\rm s},
\label{eq:v_a}
\end{align}
where
\begin{align}
c_1 = \alpha_{\rm a} \left( t_{\rm R} + T_{\rm buf} \right) - v_{\rm ini}
\label{eq:c_1}
\end{align}
and
\begin{align}
c_2 = 2 \alpha_{\rm a} \left( x_{\rm R} - X_{\rm buf} - x_{\rm a0} \right) - v_{\rm ini}^2.
\label{eq:c_2}
\end{align}
\subsection{\label{subsec:SD}Support driving (SD)}
According to Wang~\cite{Wang2018}, each SV increases its time gap parameter $T$ as an SD action to suppress the perturbations caused by the JAD actions. The details of the SD are as follows. At the initiation of a run, each SV starts increasing its $T$ from $1\,{\rm s}$ to $T_{\rm large}\,[{\rm s}]$ at an increasing rate $\beta$. Subsequently, it maintains its $T$ at $T_{\rm large}$ until the end of the run. Accordingly, its $T$ is represented as
\begin{align}
  T =
  \begin{dcases}
    1 + \beta t,   & \mbox{if $0 \le t \le \dfrac{T_{\rm large} - 1}{\beta}$,} \\
    T_{\rm large}, & \mbox{otherwise.}
  \end{dcases}
\label{eq:T_SD}
\end{align}
We define the ``supported state (Std-state)" as the equilibrium state of platoon P in which all the vehicles are traveling at velocity $v_{\rm a}$, and all the SVs maintain their time gap parameter at $T = T_{\rm large}$.

Because SVs maintain their extended time gap parameter until the end of a run, which may increase their and the subsequent vehicles' travel time, we introduce time-gap reversion mechanism, which gradually returns the extended time gap to its original value~\cite{Wang2018}. In a run without reversion, the time gap parameter is determined by Eq.~(\ref{eq:T_SD}) throughout the experiment. In a run with this mechanism, the transition begins when two conditions are satisfied simultaneously for the first time: $t \ge t_{\rm a2}$ and $v_{i_{{\rm st}, m}} (t) \ge v_{\rm cr}$. At this moment, denoted by $t = t_m^{\prime}$, vehicle $i_{{\rm st}, m}$ begins reducing its time gap $T$ from the extended value $T_{\rm large}$ back to its baseline $T = 1\,{\rm s}$ at a decreasing rate $\beta$. Here, $v_{\rm cr}\,[{\rm m}/{\rm s}]$ is the critical velocity, as defined in Sec.~\ref{subsec:string_stability}. Hence, its time gap parameter is defined by
\begin{align}
  T =
  \begin{dcases}
    1 + \beta t,   & \mbox{if $0 \le t \le \dfrac{T_{\rm large} - 1}{\beta}$,} \\
    T_{\rm large}, & \mbox{if $\dfrac{T_{\rm large} - 1}{\beta} \le t \le t_m^{\prime}$,} \\
    T_{\rm large} - \beta \left( t - t_m^{\prime} \right), & \mbox{if $t_m^{\prime} \le t \le t_m^{\prime} + \dfrac{T_{\rm large} - 1}{\beta}$,} \\
    1,             & \mbox{otherwise.}
  \end{dcases}
\label{eq:T_SD_reversion}
\end{align}
Note that $(T_{\rm large} - 1)/\beta < t_m^{\prime}$ for each SV in our settings.
\subsection{\label{subsec:string_stability}String stability}
The growth or decay of a perturbation propagating through platoon P is determined by the head-to-tail string stability~\cite{Ge2014, Wang2018, Monteil2019, Montanino2021a, Feng2019}. Let us consider platoon P in an equilibrium state with equilibrium velocity $v_{\rm e}$ and the corresponding equilibrium gap $g_{\rm e}(T_i, v_{\rm e})$ for vehicle $i$ in this platoon. Parameter $T_i \in \{ T, T_{\rm large} \}$ is vehicle $i$'s time gap parameter: $T_i = T$ for non-SVs ($i \notin A(N_{\rm st})$) and $T_i = T_{\rm large}$ for SVs ($i \in A(N_{\rm st})$). Notably, the equilibrium gap is represented as $g_{\rm e}(T_i, v_{\rm e})$ and not $g_{\rm e}(v_{\rm e})$ to emphasize that $g_{\rm e}$ depends on $T_i$.
Let us consider small perturbations $e_{{\rm g}, i}(t)$, $e_{{\rm v}, i}(t)$, and $e_{{\rm v}, i - 1}(t)$ are added to vehicle $i$'s equilibrium gap, vehicle $i$'s equilibrium velocity, and vehicle $i-1$'s equilibrium velocity, respectively. Taylor expansion of vehicle $i$'s acceleration function to linear terms around the equilibrium state is given by
\begin{align}
&f\left( g_{\rm e}(T_i, v_{\rm e}) + e_{{\rm g}, i}(t), v_{\rm e} + e_{{\rm v}, i}(t), v_{\rm e} + e_{{\rm v}, i - 1}(t) \right) \nonumber \\
&= f_{g_i}(T_i, v_{\rm e}) e_{{\rm g}, i}(t) + f_{v_i} (T_i, v_{\rm e}) e_{{\rm v}, i}(t) + f_{v_{i - 1}} (T_i, v_{\rm e}) e_{{\rm v}, i - 1}(t),
\label{eq:f_Taylor_exp}
\end{align}
where the three partial differential coefficients $f_{g_i}(T_i, v_{\rm e})$, $f_{v_i} (T_i, v_{\rm e})$, and $f_{v_{i - 1}}(T_i, v_{\rm e})$ are defined as
\begin{align}
f_{g_i}(T_i, v_{\rm e}) \coloneqq \dfrac{\partial f(g_{\rm e}(T_i, v_{\rm e}), v_{\rm e}, v_{\rm e})}{\partial g_i}
= \dfrac{ 2 a }{ g_0 + T_i v_{\rm e} } \left\{ 1 - \left( \dfrac{ v_{\rm e} }{ v_0 } \right)^{\delta} \right\}^{3/2},
\label{eq:f_g_i}
\end{align}
\begin{align}
f_{v_i} (T_i, v_{\rm e}) &\coloneqq \dfrac{\partial f(g_{\rm e}(T_i, v_{\rm e}), v_{\rm e}, v_{\rm e})}{\partial v_i} \nonumber \\
&= -a \left[ \dfrac{\delta v_{\rm e}^{\delta - 1}}{v_0^{\delta}} + \dfrac{ \left\{ 1 - \left( \dfrac{ v_{\rm e} }{ v_0 } \right)^{\delta} \right\} \left( 2 T_i + \dfrac{v_{\rm e}}{\sqrt{ab}} \right) }{ g_0 + T_i v_{\rm e} } \right],
\label{eq:f_v_i}
\end{align}
and
\begin{align}
f_{v_{i - 1}} (T_i, v_{\rm e}) \coloneqq \dfrac{\partial f(g_{\rm e}(T_i, v_{\rm e}), v_{\rm e}, v_{\rm e})}{\partial v_{i-1}}
= \dfrac{ v_{\rm e} }{ g_0 + T_i v_{\rm e} } \left\{ 1 - \left( \dfrac{ v_{\rm e} }{ v_0 } \right)^{\delta} \right\} \sqrt{ \dfrac{a}{b} },
\label{eq:f_v_i_minus_1}
\end{align}
respectively. The rightmost sides of Eqs.~(\ref{eq:f_g_i})--(\ref{eq:f_v_i_minus_1}) are derived from Eqs.~(\ref{eq:accel_IDM})--(\ref{eq:g_e_v_IDM}).
The perturbations have the following relationship:
\begin{align}
\dot{e}_{{\rm g}, i}(t)
&= \dfrac{ {\rm d} }{{\rm d} t} \left[ \left\{ x_{i - 1}(t) - x_i(t) - d \right\} - g_{\rm e}(T_i, v_{\rm e}) \right] \nonumber \\
&=  v_{i - 1}(t) - v_i(t) \nonumber \\
&= \left\{ v_{\rm e} + e_{{\rm v}, i - 1}(t) \right\} - \left\{ v_{\rm e} + e_{{\rm v}, i}(t) \right\} \nonumber \\
&= e_{{\rm v}, i - 1}(t) - e_{{\rm v}, i}(t).
\label{eq:e_g_i_t_diff}
\end{align}
Hence, the time derivative of Eq.~(\ref{eq:f_Taylor_exp}) is given by
\begin{align}
\ddot{e}_{{\rm v}, i}(t) &= f_{g_i}(T_i, v_{\rm e}) \left\{ e_{{\rm v}, i - 1}(t) - e_{{\rm v}, i}(t) \right\} + f_{v_i} (T_i, v_{\rm e}) \dot{e}_{{\rm v}, i}(t) \nonumber \\
&+ f_{v_{i - 1}} (T_i, v_{\rm e}) \dot{e}_{{\rm v}, i - 1}(t).
\label{eq:f_Taylor_exp_diff}
\end{align}
Laplace transforming Eq.~(\ref{eq:f_Taylor_exp_diff}) yields the velocity-error transfer function between vehicles $i-1$ and $i$, denoted by $G(s, T_i, v_{\rm e})$~\cite{Wang2018, Monteil2019}:
\begin{align}
G \left( s, T_i, v_{\rm e} \right) \coloneqq \dfrac{ E_{{\rm v}, i}(s) }{ E_{{\rm v}, i - 1}(s) }
= \dfrac{ f_{v_{i - 1}}(T_i, v_{\rm e}) s + f_{g_i}(T_i, v_{\rm e}) }{ s^2 - f_{v_i} (T_i, v_{\rm e}) s + f_{g_i}(T_i, v_{\rm e}) },
\label{eq:G}
\end{align}
where $s$ is the complex frequency-domain parameter, and $E_{{\rm v}, i}(s)$ and $E_{{\rm v}, i - 1}(s)$ are the Laplace transforms of $e_{{\rm v}, i}(t)$ and $e_{{\rm v}, i - 1}(t)$, respectively.
Next, we proceed to the representation of the gap-error transfer function. Laplace transforming Eq.~(\ref{eq:e_g_i_t_diff}) yields~\cite{Monteil2019}
\begin{align}
\dfrac{ E_{{\rm g}, i}(s) }{ E_{{\rm v}, i}(s) } = \dfrac{ 1 / G \left( s, T_i, v_{\rm e} \right) - 1 }{ s },
\label{eq:E_g_i_over_E_v_i}
\end{align}
where $E_{{\rm g}, i}(s)$ is the Laplace transform of $e_{{\rm g}, i}(t)$.
The gap-error transfer function between vehicles $i-1$ and $i$, denoted by $H^{\prime}(s, T_i, v_{\rm e})$, is given by
\begin{align}
H^{\prime} \left( s, T_i, v_{\rm e} \right) &\coloneqq \dfrac{ E_{{\rm g}, i}(s) }{ E_{{\rm g}, i - 1}(s) } \nonumber \\
&= \dfrac{ E_{{\rm g}, i}(s) }{ E_{{\rm v}, i}(s) } \dfrac{ E_{{\rm v}, i - 1}(s) }{ E_{{\rm g}, i - 1}(s) } \dfrac{ E_{{\rm v}, i}(s) }{ E_{{\rm v}, i - 1}(s) } \nonumber \\
&= \dfrac{ 1 / G \left( s, T_i, v_{\rm e} \right) - 1}{ 1 / G \left( s, T_{i - 1}, v_{\rm e} \right) - 1} G \left( s, T_i, v_{\rm e} \right).
\label{eq:H_prime}
\end{align}
The head-to-tail gap-error transfer function of this platoon, i.e., the gap-error transfer function from the first follower of this platoon (vehicle $i_{\rm JAD} + 2$) to the last follower (vehicle $N-1$), is represented as
\begin{align}
H \left( s, T, T_{\rm large}, A(N_{\rm st}), v_{\rm e} \right)
&= \prod_{i = i_{\rm JAD} + 2}^{N - 1} H^{\prime} \left( s, T_i, v_{\rm e} \right) \nonumber \\
&= \dfrac{ 1 / G \left( s, T_{N - 1}, v_{\rm e} \right) - 1 }{ 1 / G \left( s, T_{i_{\rm JAD} + 1}, v_{\rm e} \right) - 1 } \prod_{i = i_{\rm JAD} + 2}^{N - 1} G \left( s, T_i, v_{\rm e} \right).
\label{eq:H_1}
\end{align}
As vehicles $i_{\rm JAD} + 1$ and $N - 1$ are homogeneous in this study, the head-to-tail gap-error transfer function is simplified as follows:
\begin{align}
H \left( s, T, T_{\rm large}, A(N_{\rm st}), v_{\rm e} \right) &=  \prod_{i = i_{\rm JAD} + 2}^{N - 1} G \left( s, T_i, v_{\rm e} \right) \nonumber \\
&= \left\{ \prod_{ \substack{ i = i_{\rm JAD} + 2 \\ i \notin A(N_{\rm st}) } }^{N - 1} G \left( s, T_i, v_{\rm e} \right) \right\} \left\{ \prod_{ \substack{ i = i_{\rm JAD} + 2 \\ i \in A(N_{\rm st}) } }^{N - 1} G \left( s, T_i, v_{\rm e} \right) \right\} \nonumber \\
&= \left\{ G \left( s, T, v_{\rm e} \right) \right\}^{N - i_{\rm JAD} - k - 2} \left\{ G \left( s, T_{\rm large}, v_{\rm e} \right) \right\}^{k}.
\label{eq:H_2}
\end{align}
If, and only if, $H(s, T, T_{\rm large}, A(N_{\rm st}), v_{\rm e})$ with $s = j \omega$, where $j$ is the purely imaginary number and $\omega\,[{\rm rad}/{\rm s}]$ is the frequency of perturbation, satisfies the condition:
\begin{align}
\left| H \left( j \omega, T, T_{\rm large}, A(N_{\rm st}), v_{\rm e} \right) \right|
&= \sqrt{ \left[ \dfrac{ \left\{ f_{v_{i - 1}}(T, v_{\rm e}) \omega \right\}^2 + \left\{ f_{g_i}(T, v_{\rm e}) \right\}^2  }{ \left\{ f_{g_i}(T, v_{\rm e}) - \omega^2 \right\}^2 + \left\{ f_{v_i} (T, v_{\rm e}) \omega \right\}^2 } \right]^{N - i_{\rm JAD} - k - 2} } \nonumber \\
&\times \sqrt{ \left[ \dfrac{ \left\{ f_{v_{i - 1}}(T_{\rm large}, v_{\rm e}) \omega \right\}^2 + \left\{ f_{g_i}(T_{\rm large}, v_{\rm e}) \right\}^2  }{ \left\{ f_{g_i}(T_{\rm large}, v_{\rm e}) - \omega^2 \right\}^2 + \left\{ f_{v_i} (T_{\rm large}, v_{\rm e}) \omega \right\}^2 } \right]^k } \nonumber \\
&\le 1 \quad\text{for any $\omega \ge 0$},
\label{eq:H_norm_le_1}
\end{align}
this platoon is head-to-tail string stable~\cite{Wang2018, Monteil2019, Montanino2021a}. When Eq.~(\ref{eq:H_norm_le_1}) is satisfied, an infinitesimal perturbation caused by vehicle $i_{\rm JAD}$ decays or becomes stable as it propagates upstream of this platoon. However, if Eq.~(\ref{eq:H_norm_le_1}) is not satisfied, i.e., if
\begin{align}
\left| H \left( j \omega, T, T_{\rm large}, A(N_{\rm st}), v_{\rm e} \right) \right| > 1 \quad\text{for some $\omega \ge 0$},
\label{eq:H_norm_gt_1}
\end{align}
this platoon is head-to-tail string unstable. When Eq.~(\ref{eq:H_norm_gt_1}) is satisfied, an infinitesimal perturbation containing a frequency component of such $\omega$ is amplified as it propagates upstream of this platoon. To restrict secondary shock waves, we should maintain this platoon at a head-to-tail string-stable state. In other words, Eq.~(\ref{eq:H_norm_le_1}) should be satisfied by appropriately setting $N_{\rm st}$ and $T_{\rm large}$. Next, we consider a simple case in which the SVs are absent (i.e., $A(N_{\rm st}) = \varnothing$). In this case, Eq.~(\ref{eq:H_norm_le_1}) is simplified to~\cite{Wilson2008, Monteil2019, Montanino2021a}
\begin{align}
&\left| G \left( j \omega, T, v_{\rm e} \right) \right| \le 1 \quad\text{for any $\omega \ge 0$} \nonumber \\
&\iff \left\{ f_{v_i} (T, v_{\rm e}) \right\}^2 - \left\{ f_{v_{i - 1}}(T, v_{\rm e}) \right\}^2 - 2 f_{g_i}(T, v_{\rm e}) \ge 0.
\label{eq:G_abs_le_1}
\end{align}
In the parameter settings, Eq.~(\ref{eq:G_abs_le_1}) is satisfied if, and only if, $v_{\rm e} \ge v_{\rm cr} = 20.13\,{\rm m}/{\rm s}$, where $v_{\rm cr}$ is the critical velocity~\cite{Nishi2020}.

Note that head-to-tail string stability is also called ``weak'' string stability in some literature (e.g., Refs.~\cite{Wang2018, Monteil2019, Montanino2021a, Montanino2021b}). However, weak string stability is defined differently from head-to-tail string stability in other literature (e.g., Ref.~\cite{Feng2019}). This study uses the term head-to-tail string stability, and not weak string stability to prevent misinterpretation. See a recent review~\cite{Feng2019} for readers interested in various definitions of string stabilities.
\subsection{\label{subsec:traffic_scenarios}Traffic scenarios}
We used the following five traffic scenarios:
\begin{itemize}
\item In the no-control scenario, neither JAD nor SD is performed. In this scenario, the target shock wave caused by the perturbation of vehicle 0 propagates to vehicle $N-1$.
\item In the JAD-only scenario, only JAD is performed to resolve the target shock wave. SD is not performed in this scenario.
\item In the JAD-Support scenario, both JAD and SD are performed. After each SV increases its $T$ from $T = 1\,{\rm s}$ to $T = T_{\rm large}$, it maintains its $T$ at $T = T_{\rm large}$ until the end of the run.
\item In the JAD-SupportR scenario, both JAD and SD are performed. In contrast to the JAD-Support scenario, after the SVs have increased their time gap $T$ from $T = 1\,{\rm s}$ to $T = T_{\rm large}$, they return it to $T = 1\,{\rm s}$.
\item In the Support-only scenario, JAD is not performed, but SD, as in the JAD-Support scenario, is performed.
\end{itemize}
\subsection{\label{subsec:performance_indices}Performance indices}
We used the following three performance indices to investigate the performance of JAD and SD: total travel time $T_{\rm total}\,[{\rm s}]$, total fuel consumption $F_{\rm total}\,[{\rm kg}]$, and inverse time-to-collision $T_{\rm TC, inv}$ as a typical surrogate measure of total collision risk. The evaluation of each index for each vehicle $i$ starts at initial time $t = 0\,{\rm s}$ and ends at $t = t_{{\rm end}, i}$. Time $t_{{\rm end}, i}\,[{\rm s}]$ is the time at which vehicle $i$ reaches the target position of measurement $x_{\rm end} = 20000\,{\rm m}$.

Total travel time $T_{\rm total}$ is the time elapsed by each vehicle to travel from its initial position to $x_{\rm end}$, summed over all vehicles.
\begin{align}
T_{\rm total} = \sum_{i = 0}^{N - 1} t_{{\rm end}, i}.
\label{eq:T_total}
\end{align}

Total fuel consumption $F_{\rm total}$ is the amount of fuel consumed by each vehicle from initial time $t = 0\,{\rm s}$ to $t = t_{{\rm end}, i}$, summed over all vehicles.
\begin{align}
F_{\rm total} = \dfrac{1}{1000} \sum_{i = 0}^{N - 1} \int_{0}^{ t_{{\rm end}, i} }  F_{\rm R} (a_i (t), v_i (t)) {\rm d}t,
\label{eq:F_total}
\end{align}
where $F_{\rm R} (a, v)\,[{\rm g}/{\rm s}]$ is the instantaneous fuel consumption rate of the focal vehicle with acceleration $a\,[{\rm m}/{\rm s}^2]$ and velocity $v\,[{\rm m}/{\rm s}]$. Rate $F_{\rm R} (a, v)$ is estimated by the emissions from traffic (EMIT)~\cite{Cappiello2002} as a typical instantaneous fuel consumption model. In this estimation, $F_{\rm R} (a, v)$ is divided into the following two cases on the basis of the total tractive power requirement at the wheels $P_{\rm t} (a, v)\,[{\rm kW}]$:
\begin{align}
  F_{\rm R} (a, v) =
  \begin{dcases}
    \alpha_{\rm FR} + \beta_{\rm FR} v + \gamma_{\rm FR} v^2 + \delta_{\rm FR} v^3 + \zeta_{\rm FR} av, & \text{if $P_{\rm t} (a, v) > 0$,} \\
    \alpha_{\rm FR}^{\prime}, & \text{if $P_{\rm t} (a, v) = 0$,} \\
  \end{dcases}
\label{eq:F_R}
\end{align}
where
$\alpha_{\rm FR}\,[{\rm g}/{\rm s}]$,
$\beta_{\rm FR}\,[{\rm g}/{\rm m}]$,
$\gamma_{\rm FR}\,[{\rm g} \cdot {\rm s}/{\rm m}^2]$, 
$\delta_{\rm FR}\,[{\rm g} \cdot {\rm s}^2/{\rm m}^3]$,
$\zeta_{\rm FR}\,[{\rm g} \cdot {\rm s}^2/{\rm m}^2]$, and
$\alpha_{\rm FR}^{\prime}\,[{\rm g}/{\rm s}]$ are coefficients. Power $P_{\rm t} (a, v)$ is represented as
\begin{align}
P_{\rm t} (a, v) = \max \left\{ 0, A v + B v^2 + C v^3 + M av \right\},
\label{eq:P_t}
\end{align}
where $A\,[{\rm kW} \cdot {\rm s}/{\rm m}]$ denotes the rolling resistance coefficient, $B\,[{\rm kW} \cdot {\rm s}^2/{\rm m}^2]$ denotes the velocity correction coefficient to the rolling resistance, $C\,[{\rm kW} \cdot {\rm s}^3/{\rm m}^3]$ denotes the air-drag resistance coefficient, and $M\,[10^3\,{\rm kg}]$ denotes the mass of the vehicle. Following the approach of Cappiello et al.~\cite{Cappiello2002}, the EMIT parameters are set for vehicle category 9 (i.e., light-duty vehicles with cumulative mileages longer than 50000 miles and a high power-to-weight ratio in Tier 1 emission standards)~\cite{Barth2000}, as listed in Table~\ref{table:param}.

To define inverse time-to-collision $T_{\rm TC, inv}$, we first define the time-to-collision~\cite{Hayward1972} for vehicle $i$ at time $t$, denoted by $T_{{\rm TC}, i} (t)\,[{\rm s}]$, as the time remaining before vehicle $i$ collides with vehicle $i - 1$. It is expressed as
\begin{align}
  T_{{\rm TC}, i} (t) =
  \begin{dcases}
    \dfrac{g_i (t)}{\Delta v_i (t)} = \dfrac{x_{i - 1} (t) - x_i (t) - d}{v_i (t) - v_{i - 1} (t)}, & \text{if $v_i (t) > v_{i-1} (t)$,} \\
    \infty, & \text{if $v_i (t) \le v_{i-1} (t)$.} \\
  \end{dcases}
\label{eq:TTC_i_t}
\end{align}
The inverse time-to-collision for vehicle $i$ at time $t$, denoted by $T_{{\rm TC, inv}, i} (t)\,[{\rm s}^{-1}]$, is defined as the inverse of $T_{{\rm TC}, i} (t)$~\cite{Kiefer2005}.
\begin{align}
T_{{\rm TC, inv}, i} (t)
&= \dfrac{ 1 }{ T_{{\rm TC}, i} (t) } \nonumber \\
&= \begin{dcases}
    \dfrac{ \Delta v_i (t) }{ g_i (t) } = \dfrac{ v_i (t) - v_{i - 1} (t) }{ x_{i - 1} (t) - x_i (t) - d }, & \text{if $v_i (t) > v_{i-1} (t)$,} \\
    0, & \text{if $v_i (t) \le v_{i-1} (t)$.} \\
   \end{dcases}
\label{eq:TTC_inv_i_t}
\end{align}
As a surrogate measure of the total collision risk of the entire traffic flow within a simulation run, inverse time-to-collision $T_{\rm TC, inv}$ is defined as the total amount of $T_{{\rm TC, inv}, i} (t)$ from $t = 0\,{\rm s}$ to $t = t_{{\rm end}, i}$, summed over all vehicles except for the leading vehicle.
\begin{align}
T_{\rm TC, inv} = \sum_{i = 1}^{N - 1} \int_{0}^{ t_{{\rm end}, i} } T_{{\rm TC, inv}, i} (t) {\rm d}t.
\label{eq:TTC_inv}
\end{align}
A low (high) value of $T_{\rm TC, inv}$ denotes the collision risk of the system is low (high) within that run.
\section{\label{sec:results}Results}
Here, we set the parameter values (Table~\ref{table:param}) and conducted the following numerical calculations and simulations.
\subsection{\label{subsec:gain}Gain characteristics}
\begin{figure}[tbp]
\centering
\includegraphics[width=\hsize]{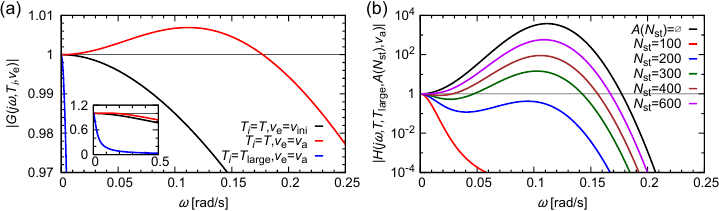}
\caption{Gain characteristics. (a) Velocity-error transfer function. (b) Head-to-tail gap-error transfer function of platoon P in the supported state (Std-state).
The thin black horizontal lines represent the vertical value of 1 as visual guides.
}
\label{fig:gain}
\end{figure}
\begin{figure}[tbp]
\centering
\includegraphics[width=\hsize]{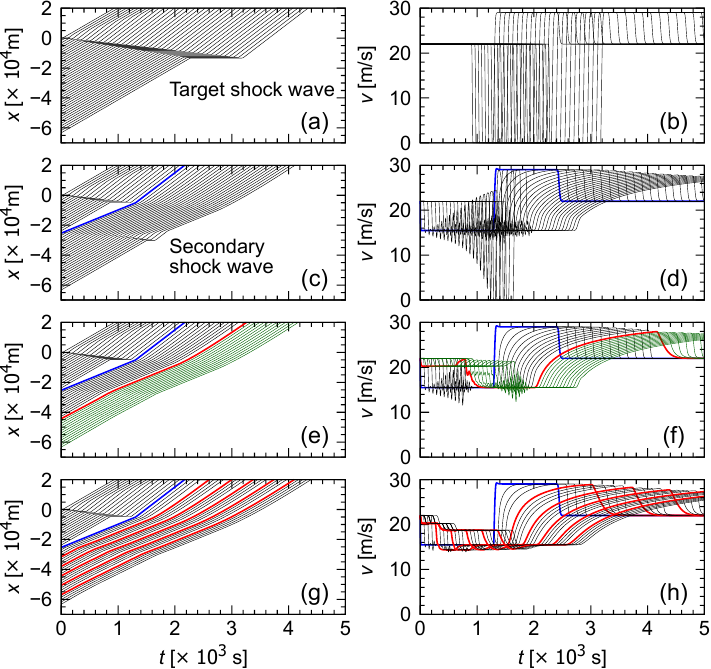}
\caption{(Left) Space-time and (Right) velocity-time diagrams. (a)--(b) No-control scenario. (c)--(d) JAD-only scenario. (e)--(f) JAD-Support scenario. $N_{\rm st} = 600$. (g)--(h) JAD-Support scenario. $N_{\rm st} = 200$. Space-time diagrams display vehicles 0, 50, 100, \ldots, 1950 and 1999. Velocity-time diagrams display vehicles 800, 850, 900, \ldots, 1950 and 1999. The absorbing vehicle (vehicle 800) and the support vehicles (SVs) are depicted by the blue and red lines, respectively, as visual guides. The vehicles upstream of the SV are depicted by the green lines in (e)--(f).}
\label{fig:xt_vt_3scenarios}
\end{figure}
Fig.~\ref{fig:gain}(a) depicts the gain characteristics of the velocity-error transfer function, i.e., the relation between $|G(j \omega, T_i, v_{\rm e})|$ and frequency $\omega$ for three cases: (i) $T_i = T = 1\,{\rm s}$ and $v_{\rm e} = v_{\rm ini} = 22\,{\rm m}/{\rm s}$, (ii) $T_i = T = 1\,{\rm s}$ and $v_{\rm e} = v_{\rm a} = 15.48\,{\rm m}/{\rm s}$, and (iii) $T_i = T_{\rm large} = 50\,{\rm s}$ and $v_{\rm e} = v_{\rm a} = 15.48\,{\rm m}/{\rm s}$. In the first case, $|G(j \omega, T, v_{\rm ini})| \le 1$ for any $\omega \ge 0\,{\rm rad}/{\rm s}$. Therefore, the initial traffic flow is head-to-tail string stable. In the second case, $|G(j \omega, T, v_{\rm a})| > 1$ for approximately $0\,{\rm rad}/{\rm s} < \omega < 0.18\,{\rm rad}/{\rm s}$. Therefore, a platoon P running at an equilibrium velocity $v_{\rm e} = v_{\rm a}$ without SD is head-to-tail string unstable. Hence, without SD, even an infinitesimal perturbation with frequency components of approximately $0\,{\rm rad}/{\rm s} < \omega < 0.18\,{\rm rad}/{\rm s}$ is amplified in platoon P. In the third case, $|G(j \omega, T_{\rm large}, v_{\rm a})| \le 1$ for any $\omega \ge 0\,{\rm rad}/{\rm s}$. Therefore, if all vehicles in platoon P travel at an equilibrium velocity $v_{\rm e} = v_{\rm a}$ and have a time gap $T_i = T_{\rm large}$, an infinitesimal perturbation decays in platoon P.

Fig.~\ref{fig:gain}(b) illustrates the gain characteristics of $H(j \omega, T, T_{\rm large}, A(N_{\rm st}), v_{\rm a})$, i.e., the relation between $|H(j \omega, T, T_{\rm large}, A(N_{\rm st}), v_{\rm a})|$ and $\omega$ for two cases in terms of the number of SVs: (i) $A(N_{\rm st}) = \varnothing$ (i.e., no SVs) and (ii) $N_{\rm st} \in \left\{ 100, 200, 300, 400, 600 \right\}$, where $N_{\rm st} = 100, 200, 300, 400$, and 600 correspond to the number of SVs $k = 11, 5, 3, 2$, and 1, respectively (see Fig.~\ref{fig:k_t_total_f_total_ittc}(a)). In the first case, $|H(j \omega, T, T_{\rm large}, \varnothing, v_{\rm a})| > 1$ for approximately $0\,{\rm rad}/{\rm s} < \omega < 0.18\,{\rm rad}/{\rm s}$. Therefore, platoon P in the Std-state is head-to-tail string unstable. An infinitesimal perturbation with frequency components of such $\omega$ evolves in platoon P. In the second case, when $N_{\rm st} \in \left\{ 100, 200 \right\}$, $|H(j \omega, T, T_{\rm large}, A(N_{\rm st}), v_{\rm a})| \le 1$ for any $\omega \ge 0\,{\rm rad}/{\rm s}$. Hence, platoon P in the Std-state is head-to-tail string stable. An infinitesimal perturbation decays in platoon P. By contrast, the gain characteristics for $N_{\rm st} \in \left\{ 300, 400, 600 \right\}$ has a $\omega$ region in which $|H(j \omega, T, T_{\rm large}, A(N_{\rm st}), v_{\rm a})| > 1$. Therefore, platoon P in the Std-state is head-to-tail string unstable. An infinitesimal perturbation containing the frequency components of such $\omega$ is amplified in platoon P. Accordingly, $N_{\rm st} \le 200$ is necessary for guaranteeing the theoretical restriction of the growth of the absorbing vehicle's perturbation.
\subsection{\label{subsec:baseline_scenarios}Baseline scenarios}
This subsection describes the results of the baseline scenarios, i.e., the no-control and JAD-only scenarios. Figs.~\ref{fig:xt_vt_3scenarios}(a) and (b) show the space-time diagram for the entire vehicle platoon and the velocity-time diagram for the absorbing vehicle and platoon P, respectively, in the no-control scenario. The target shock wave was caused by the perturbation of vehicle 0 and expanded in length as it propagated upstream, as reported in Ref.~\cite{Nishi2020}.
Figs.~\ref{fig:xt_vt_3scenarios}(c) and (d) illustrate these diagrams for the JAD-only scenario. The absorbing vehicle eliminated the target shock wave via its JAD action. However, the perturbation caused by its slow-in action evolved into the secondary shock wave, as reported in Ref.~\cite{Nishi2020}. The minimum velocity of vehicles in platoon P monotonically decreased to zero with respect to vehicle ID owing to the occurrence and propagation of the secondary shock wave.

Figs.~\ref{fig:k_t_total_f_total_ittc}(b)--(d) depict $T_{\rm total}$, $F_{\rm total}$, and $T_{\rm TC, inv}$ as functions of $N_{\rm st}$, respectively, for the no-control and JAD-only scenarios. Compared to the no-control scenario, the JAD-only scenario reduced all the three indices $T_{\rm total}$, $F_{\rm total}$, and $T_{\rm TC, inv}$ by $9.66\times10^4\,{\rm s}$ ($1.88\%$), $596\,{\rm kg}$ ($11.3\%$), and $674$ ($11.8\%$), respectively. As the velocity of vehicles within the target shock wave became zero, eliminating the target shock wave reduced $T_{\rm total}$. Meanwhile, eliminating it also precluded the possibility of unnecessary idling of vehicles within the shock wave and unnecessary acceleration of vehicles when exiting it. Such preclusions reduced $F_{\rm total}$. Furthermore, $T_{\rm TC, inv}$ generally increases when vehicles approach the ones halted ahead. Eliminating the shock wave prevented such approaching behavior of vehicles and reduced $T_{\rm TC, inv}$. However, we speculate that the secondary shock wave observed in the JAD-only scenario hindered the reduction in $F_{\rm total}$ and $T_{\rm TC, inv}$. We also speculate that the reduction in $T_{\rm total}$ was not significantly hindered because the scale of the secondary shock wave was small compared to the scale of the target shock wave, as shown in Figs.~\ref{fig:xt_vt_3scenarios}(a) and (c).
The reduction in the three indices in the JAD-only scenario agrees with the reduction in performance indices reported in the literature that dissipated or removed a shock wave: reduction in travel time~\cite{Hegyi2005, Hegyi2008, Wang2016JITS, Han2017TRC, He2017, Yang2019, Cicic2019, Zheng2020AAP, Han2021} (or increase in average velocity~\cite{ Liu2025}), fuel consumption~\cite{Wang2016JITS, Zheng2020AAP}, and collision risk~\cite{Zheng2020AAP}.
\subsection{\label{subsec:JAD_support_scenario}Scenario with both JAD and SD}
\begin{figure}[tbp]
\centering
\includegraphics[width=\hsize]{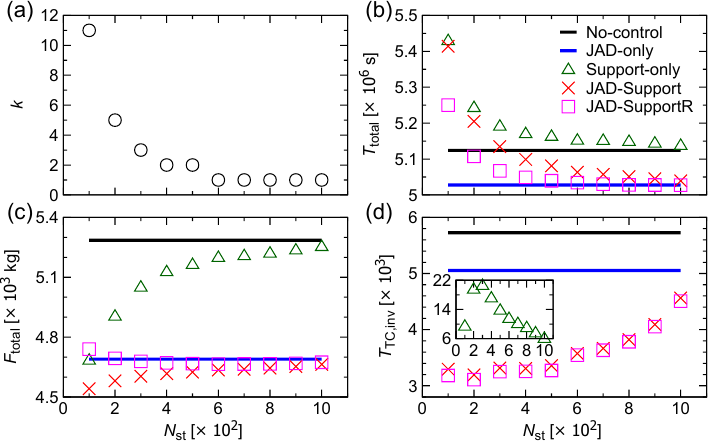}
\caption{Relations between the following output values and the difference in vehicle ID from a SV or the absorbing vehicle to the next SV, $N_{\rm st}$. (a) Number of SVs, $k$. (b) Total travel time $T_{\rm total}$. (c) Total fuel consumption $F_{\rm total}$. (d) Inverse time-to-collision $T_{\rm TC, inv}$.}
\label{fig:k_t_total_f_total_ittc}
\end{figure}
Figs.~\ref{fig:xt_vt_3scenarios}(e) and (f) depict the space-time and velocity-time diagrams, respectively, for the JAD-Support scenario with $T_{\rm large} = 50\,{\rm s}$ and $N_{\rm st} = 600$. Only one vehicle (i.e., vehicle 1400) was assigned as the SV. As shown in Fig.~\ref{fig:xt_vt_3scenarios}(e), the SV increased its time gap and restricted the perturbation caused by the absorbing vehicle from evolving into a secondary shock wave. However, as indicated in Fig.~\ref{fig:xt_vt_3scenarios}(f), because platoon P in the Std-state for $N_{\rm st} = 600$ is head-to-tail string unstable, the perturbation grew as it propagated upstream of the SV.

Figs.~\ref{fig:xt_vt_3scenarios}(g) and (h) indicate the space-time and velocity-time diagrams, respectively, for the JAD-Support scenario with $T_{\rm large} = 50\,{\rm s}$ and $N_{\rm st} = 200$. Five vehicles (vehicles 1000, 1200, 1400, 1600, and 1800) were assigned as SVs. In the case of $N_{\rm st} = 200$, the five SVs increased their time gaps and prevented secondary shock waves, as observed in Fig.~\ref{fig:xt_vt_3scenarios}(g). Because platoon P in the Std-state for $N_{\rm st} = 200$ is head-to-tail string stable, the growth of the perturbation caused by the absorbing vehicle for $N_{\rm st} = 200$ was significantly smaller than that for $N_{\rm st} = 600$, as shown in Figs.~\ref{fig:xt_vt_3scenarios}(f) and (h). In particular, the perturbation virtually disappeared upstream of the third SV.
To our knowledge, suppressing secondary shock waves by SD for a JAD scenario is successfully achieved for the first time in this study.

The four indices, $k$, $T_{\rm total}$, $F_{\rm total}$, and $T_{\rm TC, inv}$, as functions of $N_{\rm st}$ for the JAD-Support scenario are displayed in Figs.~\ref{fig:k_t_total_f_total_ittc}(a)--(d), respectively. As demonstrated in Fig.~\ref{fig:k_t_total_f_total_ittc}(a), $k$ monotonically decreased from 11 to 1 as $N_{\rm st}$ increased from 100 to 600 and was constant at 1 for $N_{\rm st}$ of 600--1000. As shown in Fig.~\ref{fig:k_t_total_f_total_ittc}(b), $T_{\rm total}$ in the JAD-Support scenario was greater than that in the JAD-only scenario over the entire range of $100 \le N_{\rm st} \le 1000$ (e.g., by $3.86\times10^5~\,{\rm s}$ ($7.68\%$) for $N_{\rm st} = 100$ and $0.12\times10^5~\,{\rm s}$ ($0.23\%$) for $N_{\rm st} = 1000$), presumably because of the following two reasons. First, the spatiotemporal scale of the secondary shock wave observed in the JAD-only scenario was small. Second, SVs increased their time gaps in the JAD-Support scenario. As $N_{\rm st}$ increased, $T_{\rm total}$ in the JAD-Support scenario decreased and approached that of the JAD-only scenario. This monotonic decrease in $T_{\rm total}$ occurred because the number of vehicles affected by the extended time gap of the SVs decreased monotonically with respect to $N_{\rm st}$. Compared to $T_{\rm total}$ in the no-control scenario, $T_{\rm total}$ in the JAD-Support scenario was greater for $100 \le N_{\rm st} \le 300$ and smaller for $400 \le N_{\rm st} \le 1000$.

As illustrated in Fig.~\ref{fig:k_t_total_f_total_ittc}(c), $F_{\rm total}$ in the JAD-Support scenario was smaller than that in the JAD-only scenario over the entire range of $100 \le N_{\rm st} \le 1000$ (e.g., by $149\,{\rm kg}$ ($3.17\%$) for $N_{\rm st} = 100$ and $26\,{\rm kg}$ ($0.55\%$) for $N_{\rm st} = 1000$), owing to the following reason. Because the SVs mitigated the perturbation or secondary shock wave caused by the absorbing vehicle, vehicles following the SVs avoided unnecessary acceleration for exiting the perturbation or secondary shock wave as well as unnecessary idling within the secondary shock wave. This avoidance reduced fuel consumption.
As $N_{\rm st}$ increased, $F_{\rm total}$ in the JAD-Support scenario increased monotonically and approached that in the JAD-only scenario. This monotonic tendency is attributed to the following reasons. As $N_{\rm st}$ increased from 100 to 600, the number of SVs, $k$, decreased from 11 to 1 (Fig.~\ref{fig:k_t_total_f_total_ittc}(a)) and $|H(j \omega, T, T_{\rm large}, A(N_{\rm st}), v_{\rm a})|$ increased (Fig.~\ref{fig:gain}(b)), i.e., platoon P became more unstable. Hence, an increase in $N_{\rm st}$ from 100 to 600 degraded the performance of the SVs in mitigating the perturbation. Even for $600 \le N_{\rm st} \le 1000$ (i.e., $k = 1$), as $N_{\rm st}$ increased from 600 to 1000, the vehicle ID of the single SV (i.e., $i_{\rm JAD} + N_{\rm st}$) increased and the perturbation was more likely to evolve into the secondary shock wave until it encountered the SV. In addition, as $N_{\rm st}$ increased from 600 to 1000, the number of vehicles following the SV (i.e., vehicles reducing their own fuel consumption owing to the SD) decreased. Accordingly, $F_{\rm total}$ in the JAD-Support scenario increased monotonically with respect to $N_{\rm st}$ for $100 \le N_{\rm st} \le 1000$.

As shown in Fig.~\ref{fig:k_t_total_f_total_ittc}(d), $T_{\rm TC, inv}$ in the JAD-Support scenario was smaller than that in the JAD-only scenario for $100 \le N_{\rm st} \le 1000$ (e.g., by $1871$ ($37.0\%$) for $N_{\rm st} = 100$ and $545$ ($10.8\%$) for $N_{\rm st} = 1000$). This relation occurred because the SVs decreased the occurrence of high-collision-risk situations, such that vehicles approached virtually or completely stopped vehicles in front of them. As $N_{\rm st}$ increased, $T_{\rm TC, inv}$ in the JAD-Support scenario tended to increase monotonically and approached that in the JAD-only scenario for the following reasons. First, as $N_{\rm st}$ increased from 100 to 600 (i.e., as $k$ decreased from 11 to 1), the capability of the SVs to stabilize platoon P deteriorated, as shown in Fig.~\ref{fig:gain}(b). Second, as $N_{\rm st}$ increased from 100 to 1000, the number of vehicles following the SVs (i.e., vehicles reducing their own collision risks owing to SD) decreased.
\subsection{\label{subsec:JAD_supportR_scenario}Scenario with both JAD and returning-time-gap-type SD}
\begin{figure}[tbp]
\centering
\includegraphics[width=\hsize]{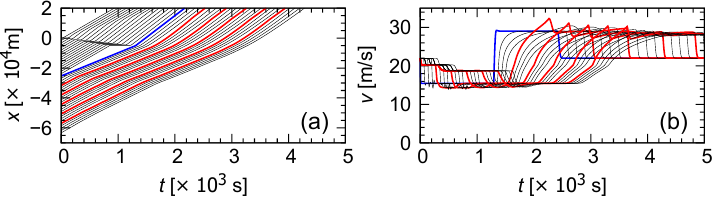}
\caption{(a) Space-time and (b) velocity-time diagrams for JAD-SupportR scenario. The line colors are set as in Figs.~\ref{fig:xt_vt_3scenarios}(g) and (h).}
\label{fig:jad_supportr_diagrams}
\end{figure}
Figs.~\ref{fig:jad_supportr_diagrams}(a) and (b) depict space-time and velocity-time diagrams, respectively, for the JAD-SupportR scenario with $N_{\rm st} = 200$. Each of the five SVs returned its time gap parameter from $T = T_{\rm large} = 50\,{\rm s}$ to the initial value $T = 1\,{\rm s}$. The SVs suppressed the perturbation caused by the slow-in action of the absorbing vehicle, and prevented the occurrence of a secondary shock wave, similarly to the JAD-Support scenario.

The indices $T_{\rm total}$, $F_{\rm total}$, and $T_{\rm TC, inv}$ as functions of $N_{\rm st}$ for the JAD-SupportR scenario are depicted in Figs.~\ref{fig:k_t_total_f_total_ittc}(b)--(d), respectively. Owing to the SVs' maneuver to return their extended time gap to the initial time gap, $T_{\rm total}$ in the JAD-SupportR scenario was smaller than that in the JAD-Support scenario for $100 \le N_{\rm st} \le 1000$ (e.g., by $1.64\times10^5\,{\rm s}$ ($3.02\%$) for $N_{\rm st} = 100$ and $0.12\times10^5,{\rm s}$ ($0.23\%$) for $N_{\rm st} = 1000$). As $N_{\rm st}$ increased from 100 to 1000, $T_{\rm total}$ in the JAD-SupportR scenario decreased monotonically and finally converged with that in the JAD-only scenario. Nevertheless, $T_{\rm total}$ in the JAD-SupportR scenario was greater than that in the JAD-only scenario, especially for $100 \le N_{\rm st} \le 300$. The relative difference in $T_{\rm total}$ between JAD-SupportR and JAD-only scenarios was $2.23\times10^5\,{\rm s}$ ($4.43$\%) for $N_{\rm st} = 100$ and $0.00\times10^5\,{\rm s}$ ($0.00$\%) for $N_{\rm st} = 1000$. Contrary to $T_{\rm total}$, $F_{\rm total}$ in the JAD-SupportR scenario was greater than that in the JAD-Support scenario for $100 \le N_{\rm st} \le 1000$ (e.g., by $199\,{\rm kg}$ ($4.38\%$) for $N_{\rm st} = 100$ and $10\,{\rm kg}$ ($0.22\%$) for $N_{\rm st} = 1000$). This is because the SVs' maneuver to return their extended time gap induced acceleration to the following vehicles, as shown in Fig.~\ref{fig:jad_supportr_diagrams}(b). The total collision risk $T_{\rm TC, inv}$ in the JAD-SupportR scenario concurred with that in the JAD-Support scenario, compared with those in the other three scenarios. Because the SVs started reverting their extended time gaps to the original value when the target shock wave was eliminated and simultaneously they recovered their own velocities (Eq.~(\ref{eq:T_SD_reversion})), their returning maneuver did not affect their own and the following vehicles' collision risk.
\subsection{\label{subsec:support_only_scenario}SD-only scenario}
\begin{figure}[tbp]
\centering
\includegraphics[width=\hsize]{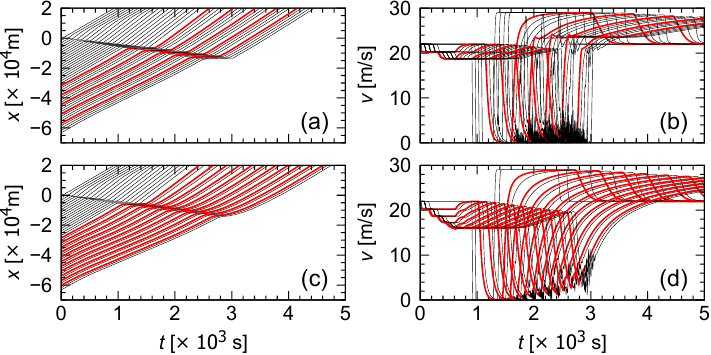}
\caption{(Left) Space-time and (Right) velocity-time diagrams for Support-only scenario. (a)--(b) $N_{\rm st} = 200$. (c)--(d) $N_{\rm st} = 100$. The SVs are depicted by the red lines as visual guides.}
\label{fig:support_only_diagrams}
\end{figure}
Figs.~\ref{fig:support_only_diagrams}(a) and (b) display the space-time and velocity-time diagrams, respectively, for the Support-only scenario with $N_{\rm st} = 200$. Figs.~\ref{fig:support_only_diagrams}(c) and (d) depict those for the Support-only scenario with $N_{\rm st} = 100$. When $N_{\rm st} = 200$, none of the five SVs could eliminate the target shock wave. Even when $N_{\rm st} = 100$, 11 SVs failed, although they mitigated it more effectively than when $N_{\rm st} = 200$. Because the target shock wave could not be eliminated, $T_{\rm total}$, $F_{\rm total}$, and $T_{\rm TC, inv}$ in the Support-only scenario (Figs.~\ref{fig:k_t_total_f_total_ittc}(b)--(d)) were greater than those in the JAD-only and JAD-Support scenarios over the entire range of $100 \le N_{\rm st} \le 1000$, except for $F_{\rm total}$ at $N_{\rm st} = 100$. For instance, $T_{\rm total}$, $F_{\rm total}$, and $T_{\rm TC, inv}$ in the Support-only scenario were greater than those in the JAD-only scenario by $4.03\times10^5\,{\rm s}$ ($8.01\%$), $-4\,{\rm kg}$ ($-0.08\%$), and $4406$ ($87.2\%$), respectively for $N_{\rm st} = 100$, and by $1.11\times10^5\,{\rm s}$ ($2.21\%$), $566\,{\rm kg}$ ($12.1\%$), and $1071$ ($21.2\%$) for $N_{\rm st} = 1000$, respectively. In particular, $T_{\rm total}$ and $T_{\rm TC, inv}$ in the Support-only scenario were the highest among those in the five scenarios. Although $T_{\rm TC, inv}$ in the Support-only scenario exhibited a single peak, as $N_{\rm st}$ increased to 1000, $T_{\rm total}$, $F_{\rm total}$, and $T_{\rm TC, inv}$ in the Support-only scenario approached those in the no-control scenario. Accordingly, JAD is necessary for fully eliminating the target shock wave over the entire range of $100 \le N_{\rm st} \le 1000$.

Contrary to the Support-only scenario in this study, Han et al.~\cite{Han2021} succeeded in dissipating a target shock wave by a hierarchical VSL control of CAVs. We speculate that such different consequences were attributed to the differences in the penetration rate of the CAVs, and the expanding rate of the target shock wave. The penetration rate of the SVs in this study was $0.55\%$ ($= 11\,{\rm vehicles} / 2000\,{\rm vehicles}$) or less, which was much smaller than that of the CAVs in Ref.~\cite{Han2021} ($5\%$). Besides, the expanding rate of the target shock wave in this study looked significantly larger than that in Ref.~\cite{Han2021}, presumably because of the difference in car-following models (this study used IDM, whereas Ref.~\cite{Han2021} used IDM+~\cite{Schakel2010}) and the parameter settings of the models. These differences led to the different consequences of the dedicated driving of the CAVs.
\section{\label{sec:conclusions}Conclusions}
The aim of this study was suppressing secondary shock waves by SD with guaranteed string stability, in a JAD scenario for clearing a target shock wave. Numerical analysis was performed on the head-to-tail string stability of platoon P in the Std-state. Numerical simulations were conducted to evaluate the effect of SD on the suppression of secondary shock waves and traffic performance indices, the effect of reverting the SV's extended time gaps, and the necessity of the combination of JAD and SD.
The following results were obtained.
\begin{enumerate}
\item[(i)] When the extended time gap $T_{\rm large}$ was fixed to $50\,{\rm s}$, platoon P in the Std-state was confirmed to be head-to-tail string stable for $N_{\rm st} \le 200$.
\item[(ii)] In the JAD-Support scenario, when $N_{\rm st} = 600$, the single SV could restrict the occurrence of secondary shock waves but amplified the perturbation caused by the absorbing vehicle's slow-in action. By contrast, when $N_{\rm st} = 200$, the five SVs successfully damped the perturbation.
\item[(iii)] Compared with the trend in the JAD-only scenario, $F_{\rm total}$ and $T_{\rm TC, inv}$ decreased but $T_{\rm total}$ increased in the JAD-Support scenario.
\item[(iv)] Compared with the trend in the JAD-Support scenario, $T_{\rm total}$ decreased, $F_{\rm total}$ increased, and $T_{\rm TC, inv}$ remained constant in the JAD-SupportR scenario.
\item[(v)] In the Support-only scenario, the target shock wave could not be fully dissolved even for $N_{\rm st} = 100$. The indices $T_{\rm total}$, $F_{\rm total}$, and $T_{\rm TC, inv}$ in this scenario were greater than those in the JAD-Support scenario.
\end{enumerate}
The numerical results demonstrate that the combination of JAD and SD has a significant impact on preventing secondary shock waves at an extremely low CAV penetration rate. Only two CAVs (one absorbing vehicle and one SV) in a platoon consisting of 2000 vehicles ($0.1\%$) were required to restrict the onset of secondary shock waves. Only six CAVs (one absorbing vehicle and five SVs, i.e., $0.3\%$) were sufficient to attenuate the perturbation upstream of the absorbing vehicle, realizing safer and more fuel-efficient traffic at the expense of total travel time. The results imply that introducing SD into JAD is suitable for improving freeway traffic under an extremely low penetration rate of CAVs.

Relaxing the assumptions stated in Sec.~\ref{subsec:assumptions}, such as those in vehicles, roads, jam prediction and estimation, and the treated secondary waves, also warrants further research. For instance, this study omitted time delays in sensors and actuators of the vehicles, and set homogeneous car-following parameters for the vehicles except the leading vehicle, absorbing vehicle during JAD actions, and SVs during SD actions. Although not a necessary and sufficient condition, Wang~\cite{Wang2018} has derived a sufficient condition for head-to-tail string stability of a heterogeneous vehicle platoon under these delays. Therefore, it will be possible to investigate the performance of SD for suppressing secondary shock waves under these delays and inter-vehicular heterogeneities.
Taking into account the variation in vehicle length, and the stochasticity of HDVs~\cite{Jiang2014, Jiang2015} as in Refs.~\cite{Liu2025, Shen2025} will be also helpful to investigate the robustness of the combination of JAD and SD.
This study focused on the secondary shock-wave problem in a JAD strategy to remove a single shock wave, as a fundamental secondary shock-wave problem. Other JAD strategies such as those for clearing multiple shock waves arising from a bottleneck~\cite{He2017, Li2024} and congestion at a sag~\cite{Nishi2022} also face secondary shock waves. Because SD is generally expected to stabilize the focal vehicle platoon, it will be possible to construct SD for such JAD strategies in order to restrict the occurrence of secondary shock waves. SD will also be applicable to moving-bottleneck-based strategies for eliminating shock waves on multiple-lane roads~\cite{Yang2019, Cicic2019, Cicic2022TRB}.
This study omitted the estimation and prediction of the target shock wave, which generally takes a finite time and contains finite errors. In future work, we will evaluate the performance of SD with an explicit estimation and prediction of the target shock wave~\cite{Netten2013, Hegyi2013} for more realistic evaluation of the combination of JAD and SD.
This study focused on the stabilization of the high-density region upstream of the absorbing vehicle, and did not address the removal of the high-density region itself. Combining SD and the spatiotemporal shrinking of the high-density region~\cite{He2025} is expected to realize more efficient and robust secondary shock-wave suppression.

As discussed in Ref.~\cite{Wang2018}, implementation of SD needs estimation of car-following parameters~\cite{Monteil2016, Ge2018, Li2020, Wang2022}. Notably, Ge and Orosz~\cite{Ge2018} have developed a real-time estimation of car-following parameters through vehicle-to-vehicle communication for mixed freeway traffic composed of connected-HDVs and CAVs; this estimation will be suitable for SD. Development in measuring high-precision vehicle trajectories~\cite{Krajewski2018, Seo2021, Gloudemans2023} is also helpful for estimation of car-following parameters. Since SD will be managed by a highway operating center, SVs will be required to be equipped with communication systems (such as vehicle-to-infrastructure~\cite{Wang2018} or vehicle-to-everything systems) to receive the commands of SD from the center. Since SD does not need many SVs (only one to five SVs for a traffic stream of 2000 vehicles), SVs are expected to be deployed by police organizations and/or freeway operating companies, similarly to the absorbing vehicle~\cite{He2025}. Implementing more realistic SD taking into account these requirements and situations will be studied as future work.
In this study, SVs started their SD actions simultaneously, and had the same settings of the SD parameters (such as the extended time gap) for simplicity. In practice, SVs need to dynamically set their SD parameters reflecting the stability of the traffic flow downstream of them. Moreover, if the performance of SD in stabilizing platoon P is turned out to be lower than expected during the operation of SD, additional SVs should be assigned to reinforce the stabilizing effect. Such dynamical design of SD for JAD strategies will be developed in future work.
Interestingly, a recent RL-based control of AVs also introduced extended time gaps of AVs to attenuate shock waves~\cite{Jang2025}. Therefore, introducing RL into SD should further improve JAD.
Conducting field experiments to confirm the robustness of the combination of JAD and SD is also an open problem.
It is known that the head-to-tail string stability condition used in this study (i.e., Eq.~(\ref{eq:H_norm_le_1})) captures the energy of the state deviations from the equilibrium state, and not their maximum amplitude because Eq.~(\ref{eq:H_norm_le_1}) is characterized by $\mathcal{L}_2$ norm, and not by $\mathcal{L}_{\infty}$ norm~\cite{Feng2019, Montanino2021b}. Since Montanino et al.~\cite{Montanino2021b} reported that $\mathcal{L}_{\infty}$ string stability captures the propagation of disturbances in vehicle platoons more appropriately than $\mathcal{L}_2$ string stability, we will investigate the effectiveness of SD in terms of $\mathcal{L}_{\infty}$ string stability in future work.

\section*{Acknowledgements}
This work was supported by JSPS KAKENHI Grant Number JP23K04287.

\section*{ORCID}
\noindent Ryosuke Nishi - \url{https://orcid.org/0000-0002-4788-673X}

\begin{appendices}
\section{\label{sec:nomenclature}Nomenclature}
The nomenclature is listed in Tables~\ref{table:nomenclature}--~\ref{table:nomenclature_continue_2}. 
\begin{table}[t]
\centering
\caption{
Nomenclature.\label{table:nomenclature}}
{\begin{tabular}{@{}lll@{}}
\toprule
Name or Symbol & Unit & Description \\
\midrule
$A$ & $[{\rm kW} \cdot {\rm s}/{\rm m}]$ & Rolling resistance coefficient  \\
$A(N_{\rm st})$ & - & Set of IDs of support vehicles for a given $N_{\rm st}$ \\
ACC & - & Adaptive-cruise control \\
AV & - & Automated vehicle \\
$a$ & $[{\rm m}/{\rm s}^2]$ & Maximum acceleration in IDM \\
$a_{i}(t)$ & $[{\rm m}/{\rm s}^2]$ & Vehicle $i$'s acceleration at time $t$ \\
absorbing vehicle & - & Vehicle that performs JAD \\
absorbing velocity & - & Absorbing vehicle's velocity in performing JAD \\
$B$ & $[{\rm kW} \cdot {\rm s}^2/{\rm m}^2]$ & Velocity correction coefficient to \\
& & the rolling resistance  \\
$b$ & $[{\rm m}/{\rm s}^2]$ & Comfortable deceleration in IDM \\
$C$ & $[{\rm kW} \cdot {\rm s}^3/{\rm m}^3]$ & Air-drag resistance coefficient \\
CAV & - & Connected and automated vehicle \\
CFC & - & Car-following control \\
CV & - & Connected vehicle \\
$d$ & $[{\rm m}]$ & Vehicle length \\
$E_{{\rm g}, i}(s)$ & - & Laplace transform of $e_{{\rm g}, i}(t)$ \\
$E_{{\rm v}, i}(s)$ & - & Laplace transform of $e_{{\rm v}, i}(t)$ \\
EMIT & - & Emissions from traffic \\
$e_{{\rm g}, i}(t)$ & [m] & A small perturbation of vehicle $i$'s gap \\
$e_{{\rm v}, i}(t)$ & $[{\rm m}/{\rm s}]$ & A small perturbation of vehicle $i$'s velocity \\
$e_{{\rm v}, i - 1}(t)$ & $[{\rm m}/{\rm s}]$ & A small perturbation of vehicle $i-1$'s velocity \\
$F_{\rm R}(a, v)$ & $[{\rm g}/{\rm s}]$ & Instantaneous fuel consumption rate \\
$F_{\rm total}$ & [kg] & Total fuel consumption \\
$f(g_i(t), v_i(t), v_{i-1}(t))$ & $[{\rm m}/{\rm s}^2]$ & Vehicle $i$'s acceleration as a function of \\
& & $g_i(t)$, $v_i(t)$, and $v_{i-1}(t)$ \\
$f_{g_i}(T_i, v_{\rm e})$ & $[{\rm s}^{-2}]$ & Partial derivative of $f$ with respect to $g_i$ \\
& & in equilibrium at $T_i$ and $v_{\rm e}$ \\
$f_{v_i} (T_i, v_{\rm e})$ & $[{\rm s}^{-1}]$ & Partial derivative of $f$ with respect to $v_i$ \\
& & in equilibrium at $T_i$ and $v_{\rm e}$ \\
$f_{v_{i - 1}} (T_i, v_{\rm e})$ & $[{\rm s}^{-1}]$ & Partial derivative of $f$ with respect to $v_{i-1}$ \\
& & in equilibrium at $T_i$ and $v_{\rm e}$ \\
$G(s, T_i, v_{\rm e})$ & - & Velocity-error transfer function between \\
& & vehicles $i-1$ and $i$ \\
$g^{*}\left( v_i(t), \Delta v_i(t) \right)$ & [m] & Desired dynamic gap in IDM \\
$g_0$ & $[{\rm m}]$ & Gap in the stopped state in IDM \\
$g_{\rm e}(T_i, v)$ & [m] & Equilibrium gap for safe time gap $T_i$ and velocity $v$ \\
$g_{\rm e}(v)$ & [m] & Equilibrium gap for velocity $v$ \\
$g_{i}(t)$ & [m] & Vehicle $i$'s gap at time $t$ \\
$H \left( s, T, T_{\rm large}, A(N_{\rm st}), v_{\rm e} \right)$ & - & Head-to-tail gap-error transfer function \\
$H^{\prime}(s, T_i, v_{\rm e})$ & - & Gap-error transfer function between \\
& & vehicles $i-1$ and $i$ \\
IDM & - & Intelligent driver model \\
$i$ & - & Vehicle identifier (ID) \\
$i_{\rm JAD}$ & - & ID of the absorbing vehicle  \\
$i_{{\rm st}, m}$ & - & ID of the $m$th support vehicle \\
JAD & - & Jam-absorption driving \\
JAD-Support scenario & - & Scenario in which both JAD and SD are performed \\
JAD-SupportR scenario & - & JAD-Support scenario with reversion \\
& & of SVs' extended time gap \\
\bottomrule
\end{tabular}}
\end{table}

\begin{table}[t]
\centering
\caption{
Nomenclature (Continued).\label{table:nomenclature_continue}}
{\begin{tabular}{@{}lll@{}}
\toprule
Name or Symbol & Unit & Description \\
\midrule
JAD-only scenario & - & Scenario in which JAD is performed \\
& & but SD is not performed \\
$j$ & - & Purely imaginary number \\
$k$ & - & Number of support vehicles \\
$M$ & $[10^3\,{\rm kg}]$ & Vehicle mass \\
MPC & - & Model predictive control \\
$N$ & - & Number of vehicles \\
$N_{\rm st}$ & - & Difference in vehicle ID from a SV or the \\
& & absorbing vehicle to the next SV \\
no-control scenario & - & Scenario in which neither JAD nor SD is performed \\
$P_{\rm t}(a, v) $ & [kW] & Total tractive power requirement at the wheels \\
platoon P & - & Platoon composed of all the vehicles \\
& & upstream of the absorbing vehicle \\
SD & - & Support driving \\
SPECIALIST & - & SPEed ControllIng ALgorIthm using Shockwave Theory \\
SV & - & Support vehicle \\
Std-state & - & Supported state, an equilibrium state of platoon P \\
& & at $T = T_{\rm large}$ for all SVs and velocity $v_{\rm a}$ \\
Support-only scenario & - & Scenario in which SD is performed \\
& & but JAD is not performed \\
$s$ & - & Complex frequency-domain parameter \\
$T$ & $[{\rm s}]$ & Safe time gap in IDM \\
$T_{{\rm TC}, i} (t)$ & [s] & Time-to-collision for vehicle $i$ at time $t$ \\
$T_{\rm TC,\,inv}$ & 1 & Inverse time-to-collision \\
$T_{{\rm TC, inv}, i} (t)$ & $[{\rm s}^{-1}]$ & Inverse time-to-collision for vehicle $i$ at time $t$ \\ 
$T_{\rm buf}$ & $[{\rm s}]$ & Time buffer in JAD  \\
$T_i$ & [s] & Vehicle $i$'s safe time gap in IDM \\
$T_{\rm large}$ & $[{\rm s}]$ & Extended time gap in SD \\
$T_{\rm total}$ & [s] & Total travel time \\
$t$ & [s] & Time \\
$t_{\rm R}$ & [s] & Time when the vehicle in front of the absorbing \\
& & vehicle exits the target shock wave \\
$t_{\rm a1}$ & [s] & Time when the absorbing vehicle's velocity becomes $v_{\rm a}$ \\
$t_{\rm a2}$ & [s] & Time when the absorbing vehicle ends running at $v_{\rm a}$ \\
$t_{{\rm end}, i}$ & [s] & Time at which vehicle $i$ reaches $x_{\rm end}$ \\
$t_m^{\prime}$ & [s] & The first time when $t \ge t_{\rm a2}$ and $v_{i_{{\rm st}, m}} (t) \ge v_{\rm cr}$ \\
VSL & - & Variable speed limit \\
$v_0$ & $[{\rm m}/{\rm s}]$ & Ideal velocity in IDM \\
$v_{\rm a}$ & $[{\rm m}/{\rm s}]$ & Absorbing velocity \\
$v_{\rm cr}$ & $[{\rm m}/{\rm s}]$ & Critical velocity in terms of linear string stability \\
$v_{\rm e}$ & $[{\rm m}/{\rm s}]$ & Equilibrium velocity \\
$v_{i}(t)$ & $[{\rm m}/{\rm s}]$ & Vehicle $i$'s velocity at time $t$ \\
$v_{\rm ini}$ & $[{\rm m}/{\rm s}]$ & Initial velocity \\
$X_{\rm buf}$ & $[{\rm m}]$ & Spatial buffer in JAD \\
$x_{\rm R}$ & [m] & Position where the vehicle in front of the absorbing \\
& & vehicle exits the target shock wave \\
$x_{\rm a0}$ & [m] & Initial position of the absorbing vehicle \\
$x_{\rm a1}$ & [m] & Position where the absorbing vehicle's velocity \\
& & becomes $v_{\rm a}$ \\
$x_{\rm a2}$ & [m] & Position where the absorbing vehicle ends running at $v_{\rm a}$ \\
\bottomrule
\end{tabular}}
\end{table}

\clearpage
\begin{table}[t]
\centering
\caption{
Nomenclature (Continued).\label{table:nomenclature_continue_2}}
{\begin{tabular}{@{}lll@{}}
\toprule
Name or Symbol & Unit & Description \\
\midrule
$x_{\rm end}$ & $[{\rm m}]$ & Measurement end position \\
$x_{i}(t)$ & [m] & Vehicle $i$'s position at time $t$ \\
$\alpha_{\rm a}$ & $[{\rm m}/{\rm s}^2]$ & Deceleration in JAD \\
$\alpha_{\rm FR}$ & $[{\rm g}/{\rm s}]$ & Coefficient in $F_{\rm R}(a, v)$ \\
$\alpha_{\rm FR}^{\prime}$ & $[{\rm g}/{\rm s}]$ & Coefficient in $F_{\rm R}(a, v)$ \\
$\beta$ & - & Increasing/decreasing rate of time gap in SD \\
$\beta_{\rm FR}$ & $[{\rm g}/{\rm m}]$ & Coefficient in $F_{\rm R}(a, v)$ \\
$\gamma_{\rm FR}$ & $[{\rm g} \cdot {\rm s}/{\rm m}^2]$ & Coefficient in $F_{\rm R}(a, v)$ \\
$\Delta t$ & $[{\rm s}]$ & Time interval in the numerical integration scheme \\
$\delta$ & - & Power coefficient in IDM \\
$\delta_{\rm FR}$ & $[{\rm g} \cdot {\rm s}^2/{\rm m}^3]$ & Coefficient in $F_{\rm R}(a, v)$ \\
$\zeta_{\rm FR}$ & $[{\rm g} \cdot {\rm s}^2/{\rm m}^2]$ & Coefficient in $F_{\rm R}(a, v)$ \\
$\omega$ & $[{\rm rad}/{\rm s}]$ & Frequency of perturbation \\
\bottomrule
\end{tabular}}
\end{table}
\end{appendices}


\begin{thebibliography}{99}

\bibitem{Schrank2024}
David Schrank, Luke Albert, Kartikeya Jha, and Bill Eisele.
\newblock 2023 urban mobility report.
\newblock The Texas A\&M Transportation Institute with cooperation from INRIX,
  2024.
\newblock
  \url{https://static.tti.tamu.edu/tti.tamu.edu/documents/mobility-report-2023.pdf}.

\bibitem{Lu2014}
Xiao-Yun Lu and Steven~E Shladover.
\newblock Review of variable speed limits and advisories: Theory, algorithms,
  and practice.
\newblock {\em Transp. Res. Rec.}, 2423(1):15--23, 2014.
\newblock \url{https://doi.org/10.3141/2423-03}.

\bibitem{Khondaker2015}
Bidoura Khondaker and Lina Kattan.
\newblock Variable speed limit: An overview.
\newblock {\em Transp. Lett.}, 7(5):264--278, 2015.
\newblock \url{https://doi.org/10.1179/1942787514Y.0000000053}.

\bibitem{Papageorgiou2002}
Markos Papageorgiou and Apostolos Kotsialos.
\newblock Freeway ramp metering: An overview.
\newblock {\em IEEE Trans. Intell. Transp. Syst.}, 3(4):271--281, 2002.
\newblock \url{https://doi.org/10.1109/TITS.2002.806803}.

\bibitem{Papageorgiou1991}
Markos Papageorgiou, Habib Hadj-Salem, and Jean-Marc Blosseville.
\newblock {ALINEA}: A local feedback control law for on-ramp metering.
\newblock {\em Transp. Res. Rec.}, 1320:58--64, 1991.

\bibitem{Hegyi2005}
Andreas Hegyi, Bart De~Schutter, and Johannes Hellendoorn.
\newblock Optimal coordination of variable speed limits to suppress shock
  waves.
\newblock {\em IEEE Trans. Intell. Transp. Syst.}, 6(1):102--112, 2005.
\newblock \url{https://doi.org/10.1109/TITS.2004.842408}.

\bibitem{Han2017TRC}
Yu~Han, Andreas Hegyi, Yufei Yuan, Serge Hoogendoorn, Markos Papageorgiou, and
  Claudio Roncoli.
\newblock Resolving freeway jam waves by discrete first-order model-based
  predictive control of variable speed limits.
\newblock {\em Transp. Res. C}, 77:405--420, 2017.
\newblock \url{https://doi.org/10.1016/j.trc.2017.02.009}.

\bibitem{Hegyi2008}
Andreas Hegyi, Serge~P Hoogendoorn, Marco Schreuder, Henk Stoelhorst, and
  Francesco Viti.
\newblock {SPECIALIST}: A dynamic speed limit control algorithm based on shock
  wave theory.
\newblock In {\em 2008 11th Int. IEEE Conf. Intell. Transp. Syst.}, pages
  827--832. IEEE, 2008.
\newblock \url{https://doi.org/10.1109/ITSC.2008.4732611}.

\bibitem{Hegyi2010}
Andreas Hegyi and Serge~P Hoogendoorn.
\newblock Dynamic speed limit control to resolve shock waves on freeways ---
  {F}ield test results of the {SPECIALIST} algorithm.
\newblock In {\em 13th Int. IEEE Conf. Intell. Transp. Syst.}, pages 519--524.
  IEEE, 2010.
\newblock \url{https://doi.org/10.1109/ITSC.2010.5624974}.

\bibitem{Carlson2010TranspSci}
Rodrigo~C. Carlson, Ioannis Papamichail, Markos Papageorgiou, and Albert
  Messmer.
\newblock {Optimal motorway traffic flow control involving variable speed
  limits and ramp metering}.
\newblock {\em Transp. Sci.}, 44(2):238--253, 2010.
\newblock \url{https://doi.org/10.1287/trsc.1090.0314}.

\bibitem{Carlson2010TRC}
Rodrigo~C. Carlson, Ioannis Papamichail, Markos Papageorgiou, and Albert
  Messmer.
\newblock {Optimal mainstream traffic flow control of large-scale motorway
  networks}.
\newblock {\em Transp. Res. C}, 18(2):193--212, 2010.
\newblock \url{https://doi.org/10.1016/j.trc.2009.05.014}.

\bibitem{Yu2021}
Haiyang Yu, Rui Jiang, Zhengbing He, Zuduo Zheng, Li~Li, Runkun Liu, and Xiqun
  Chen.
\newblock Automated vehicle-involved traffic flow studies: A survey of
  assumptions, models, speculations, and perspectives.
\newblock {\em Transp. Res. C}, 127:103101, 2021.
\newblock \url{https://doi.org/10.1016/j.trc.2021.103101}.

\bibitem{Kesting2008}
Arne Kesting, Martin Treiber, Martin Sch{\"o}nhof, and Dirk Helbing.
\newblock Adaptive cruise control design for active congestion avoidance.
\newblock {\em Transp. Res. C}, 16(6):668--683, 2008.
\newblock \url{https://doi.org/10.1016/j.trc.2007.12.004}.

\bibitem{Knorr2012}
Florian Knorr, Daniel Baselt, Michael Schreckenberg, and Martin Mauve.
\newblock Reducing traffic jams via {VANETs}.
\newblock {\em IEEE Trans. Veh. Technol.}, 61(8):3490--3498, 2012.
\newblock \url{https://doi.org/10.1109/TVT.2012.2209690}.

\bibitem{Han2021}
Yu~Han, Meng Wang, Ziang He, Zhibin Li, Hao Wang, and Pan Liu.
\newblock A linear {L}agrangian model predictive controller of macro- and
  micro- variable speed limits to eliminate freeway jam waves.
\newblock {\em Transp. Res. C}, 128:103121, 2021.
\newblock \url{https://doi.org/10.1016/j.trc.2021.103121}.

\bibitem{Li2019}
Li~Li and Xiaopeng Li.
\newblock Parsimonious trajectory design of connected automated traffic.
\newblock {\em Transp. Res. B}, 119:1--21, 2019.
\newblock \url{https://doi.org/10.1016/j.trb.2018.11.006}.

\bibitem{Beaty1998}
William~J. Beaty.
\newblock {Traffic ``experiments" and a cure for waves \& jams}, 1998.
\newblock \url{http://trafficwaves.org/trafexp.html} (Accessed: January 9,
  2026).

\bibitem{Behl2010}
Madhur Behl and Rahul Mangharam.
\newblock {Pacer cars: Real-time traffic shockwave suppression}.
\newblock In {\em 2010 31st IEEE Real-Time Syst. Symp., Work-in-Prog. Sess.}
  IEEE, 2010.
\newblock \url{https://cse.unl.edu/~rtss2008/archive/rtss2010/WIP2010/11.pdf}.

\bibitem{Washino2003}
Shoichi Washino.
\newblock Improvement of traffic flow and preservation of the environment.
\newblock {\em Bull. Tottori Univ. Environ. Stud.}, 1:61--67, 2003.
\newblock
  \url{https://www.kankyo-u.ac.jp/f/introduction/publication/bulletin/1/5.pdf}
  (in Japanese).

\bibitem{Nishi2013}
Ryosuke Nishi, Akiyasu Tomoeda, Kenichiro Shimura, and Katsuhiro Nishinari.
\newblock Theory of jam-absorption driving.
\newblock {\em Transp. Res. B}, 50:116--129, 2013.
\newblock \url{https://doi.org/10.1016/j.trb.2013.02.003}.

\bibitem{Taniguchi2015}
Yohei Taniguchi, Ryosuke Nishi, Takahiro Ezaki, and Katsuhiro Nishinari.
\newblock Jam-absorption driving with a car-following model.
\newblock {\em Phys. A}, 433:304--315, 2015.
\newblock \url{https://doi.org/10.1016/j.physa.2015.03.036}.

\bibitem{He2017}
Zhengbing He, Liang Zheng, Liying Song, and Ning Zhu.
\newblock A jam-absorption driving strategy for mitigating traffic
  oscillations.
\newblock {\em IEEE Trans. Intell. Transp. Syst.}, 18(4):802--813, 2017.
\newblock \url{https://doi.org/10.1109/TITS.2016.2587699}.

\bibitem{Han2017TRB}
Youngjun Han, Danjue Chen, and Soyoung Ahn.
\newblock {Variable speed limit control at fixed freeway bottlenecks using
  connected vehicles}.
\newblock {\em Transp. Res. B}, 98:113--134, 2017.
\newblock \url{https://doi.org/10.1016/j.trb.2016.12.013}.

\bibitem{Stern2018}
Raphael~E Stern, Shumo Cui, Maria~Laura Delle~Monache, Rahul Bhadani, Matt
  Bunting, Miles Churchill, Nathaniel Hamilton, R'mani Haulcy, Hannah Pohlmann,
  Fangyu Wu, Benedetto Piccoli, Benjamin Seibold, Jonathan Sprinkle, and
  Daniel~B Work.
\newblock Dissipation of stop-and-go waves via control of autonomous vehicles:
  Field experiments.
\newblock {\em Transp. Res. C}, 89:205--221, 2018.
\newblock \url{https://doi.org/10.1016/j.trc.2018.02.005}.

\bibitem{Ghiasi2019}
Amir Ghiasi, Xiaopeng Li, and Jiaqi Ma.
\newblock {A mixed traffic speed harmonization model with connected autonomous
  vehicles}.
\newblock {\em Transp. Res. C}, 104:210--233, 2019.
\newblock \url{https://doi.org/10.1016/j.trc.2019.05.005}.

\bibitem{Zheng2020ITJ}
Yang Zheng, Jiawei Wang, and Keqiang Li.
\newblock Smoothing traffic flow via control of autonomous vehicles.
\newblock {\em IEEE Internet Things J.}, 7(5):3882--3896, 2020.
\newblock \url{https://doi.org/10.1109/JIOT.2020.2966506}.

\bibitem{Wu2022}
Cathy Wu, Abdul~Rahman Kreidieh, Kanaad Parvate, Eugene Vinitsky, and
  Alexandre~M Bayen.
\newblock Flow: A modular learning framework for mixed autonomy traffic.
\newblock {\em IEEE Trans. Robot.}, 38(2):1270--1286, 2022.
\newblock \url{https://doi.org/10.1109/TRO.2021.3087314}.

\bibitem{Nishi2022}
Ryosuke Nishi and Takashi Watanabe.
\newblock System-size dependence of a jam-absorption driving strategy to remove
  traffic jam caused by a sag under the presence of traffic instability.
\newblock {\em Phys. A}, 600:127512, 2022.
\newblock \url{https://doi.org/10.1016/j.physa.2022.127512}.

\bibitem{He2025}
Zhengbing He, Jorge Laval, Yu~Han, Andreas Hegyi, Ryosuke Nishi, and Cathy Wu.
\newblock A review of stop-and-go traffic wave suppression strategies: Variable
  speed limit vs. jam-absorption driving.
\newblock {\em arXiv preprint arXiv:2504.11372v2}, 2025.
\newblock \url{https://doi.org/10.48550/arXiv.2504.11372}.

\bibitem{Ramadan2017}
Rabie~A Ramadan and Benjamin Seibold.
\newblock Traffic flow control and fuel consumption reduction via moving
  bottlenecks.
\newblock {\em arXiv preprint arXiv:1702.07995}, 2017.
\newblock \url{https://doi.org/10.48550/arXiv.1702.07995}.

\bibitem{Piacentini2018}
Giulia Piacentini, Paola Goatin, and Antonella Ferrara.
\newblock Traffic control via moving bottleneck of coordinated vehicles.
\newblock {\em IFAC-Pap.}, 51(9):13--18, 2018.
\newblock \url{https://doi.org/10.1016/j.ifacol.2018.07.003}.

\bibitem{Yang2019}
Hao Yang and Ken Oguchi.
\newblock Multi-lane freeway oscillation mitigation at early-stage development
  of connected vehicles.
\newblock In {\em 2019 IEEE Intell. Veh. Symp. (IV)}, pages 2072--2079. IEEE,
  2019.
\newblock \url{https://doi.org/10.1109/IVS.2019.8813867}.

\bibitem{Wang2018}
Meng Wang.
\newblock Infrastructure assisted adaptive driving to stabilise heterogeneous
  vehicle strings.
\newblock {\em Transp. Res. C}, 91:276--295, 2018.
\newblock \url{https://doi.org/10.1016/j.trc.2018.04.010}.

\bibitem{Feng2019}
Shuo Feng, Yi~Zhang, Shengbo~Eben Li, Zhong Cao, Henry~X Liu, and Li~Li.
\newblock String stability for vehicular platoon control: Definitions and
  analysis methods.
\newblock {\em Annu. Rev. Control}, 47:81--97, 2019.
\newblock \url{https://doi.org/10.1016/j.arcontrol.2019.03.001}.

\bibitem{Wang2012}
Yizhi Wang, YI~Zhang, Jianming Hu, and Li~Li.
\newblock Using variable speed limits to eliminate wide moving jams: A study
  based on three-phase traffic theory.
\newblock {\em Int. J. Mod. Phys. C}, 23(9):1250060, 2012.
\newblock \url{https://doi.org/10.1142/S012918311250060X}.

\bibitem{Wang2014}
Yizhi Wang, Jianming Hu, Li~Li, and Yi~Zhang.
\newblock Optimal coordination of variable speed limit to eliminate freeway
  wide moving jams.
\newblock {\em Int. J. Mod. Phys. C}, 25(9):1450038, 2014.
\newblock \url{https://doi.org/10.1142/S0129183114500387}.

\bibitem{Wang2016JITS}
Meng Wang, Winnie Daamen, Serge~P Hoogendoorn, and Bart {van Arem}.
\newblock Connected variable speed limits control and car-following control
  with vehicle-infrastructure communication to resolve stop-and-go waves.
\newblock {\em J. Intell. Transp. Syst.}, 20(6):559--572, 2016.
\newblock \url{https://doi.org/10.1080/15472450.2016.1157022}.

\bibitem{Nishi2020}
Ryosuke Nishi.
\newblock Theoretical conditions for restricting secondary jams in
  jam-absorption driving scenarios.
\newblock {\em Phys. A}, 542:123393, 2020.
\newblock \url{https://doi.org/10.1016/j.physa.2019.123393}.

\bibitem{Zheng2020AAP}
Yuan Zheng, Guoqiang Zhang, Ye~Li, and Zhibin Li.
\newblock Optimal jam-absorption driving strategy for mitigating rear-end
  collision risks with oscillations on freeway straight segments.
\newblock {\em Accid. Anal. Prev.}, 135:105367, 2020.
\newblock \url{https://doi.org/10.1016/j.aap.2019.105367}.

\bibitem{Li2024}
Siyu Li, Daichi Yanagisawa, and Katsuhiro Nishinari.
\newblock A jam-absorption driving system for reducing multiple moving jams by
  estimating moving jam propagation.
\newblock {\em Transp. Res. C}, 158:104394, 2024.
\newblock \url{https://doi.org/10.1016/j.trc.2023.104394}.

\bibitem{Liu2025}
Can Liu, Fangfang Zheng, Henry~X Liu, and Xiaobo Liu.
\newblock Optimizing mixed traffic flow: Longitudinal control of connected and
  automated vehicles to mitigate traffic oscillations.
\newblock {\em IEEE Trans. Intell. Transp. Syst.}, 26(3):3482--3498, 2025.
\newblock \url{https://doi.org/10.1109/TITS.2024.3522002}.

\bibitem{Cicic2019}
Mladen {\v{C}}i{\v{c}}i{\'c} and Karl~Henrik Johansson.
\newblock Stop-and-go wave dissipation using accumulated controlled moving
  bottlenecks in multi-class {CTM} framework.
\newblock In {\em 2019 IEEE 58th Conf. Decis. Control}, pages 3146--3151. IEEE,
  2019.
\newblock \url{https://doi.org/10.1109/CDC40024.2019.9029216}.

\bibitem{Cicic2022TRB}
Mladen {\v{C}}i{\v{c}}i{\'c} and Karl~Henrik Johansson.
\newblock Front-tracking transition system model for traffic state
  reconstruction, model learning, and control with application to stop-and-go
  wave dissipation.
\newblock {\em Transp. Res. B}, 166:212--236, 2022.
\newblock \url{https://doi.org/10.1016/j.trb.2022.10.008}.

\bibitem{Jiang2014}
Rui Jiang, Mao-Bin Hu, H~M Zhang, Zi-You Gao, Bin Jia, Qing-Song Wu, Bing Wang,
  and Ming Yang.
\newblock Traffic experiment reveals the nature of car-following.
\newblock {\em PLoS ONE}, 9(4):e94351, 2014.
\newblock \url{https://doi.org/10.1371/journal.pone.0094351}.

\bibitem{Jiang2015}
Rui Jiang, Mao-Bin Hu, H~M Zhang, Zi-You Gao, Bin Jia, and Qing-Song Wu.
\newblock On some experimental features of car-following behavior and how to
  model them.
\newblock {\em Transp. Res. B}, 80:338--354, 2015.
\newblock \url{https://doi.org/10.1016/j.trb.2015.08.003}.

\bibitem{Netten2013}
Bart Netten, Andreas Hegyi, Meng Wang, WJ~Schakel, Yufei Yuan, Thomas
  Schreiter, Bart {van Arem}, Coen {van Leeuwen}, and Tom Alkim.
\newblock Improving moving jam detection performance with {V2I} communication.
\newblock In {\em 20th World Cong. Intell. Transp. Syst. 2013}, pages
  5332--5341. Intelligent Transportation Society of Japan, 2013.

\bibitem{Hegyi2013}
Andreas Hegyi, Bart~D Netten, Meng Wang, W~Schakel, Thomas Schreiter, Yufei
  Yuan, Bart van Arem, and Tom Alkim.
\newblock A cooperative system based variable speed limit control algorithm
  against jam waves --- {An} extension of the {SPECIALIST} algorithm.
\newblock In {\em 16th Int. IEEE Conf. Intell. Transp. Syst. (ITSC 2013)},
  pages 973--978. IEEE, 2013.
\newblock \url{https://doi.org/10.1109/ITSC.2013.6728358}.

\bibitem{Han2015}
Youngjun Han, Danjue Chen, Soyoung Ahn, and Andreas Hegyi.
\newblock Analysis of driver response and traffic evolution under variable
  speed limit control.
\newblock {\em Transp. Res. Rec.}, 2490:1--10, 2015.
\newblock \url{https://doi.org/10.3141/2490-01}.

\bibitem{Wang2023}
Yizhou Wang and Peter~J Jin.
\newblock Model predictive control policy design, solutions, and stability
  analysis for longitudinal vehicle control considering shockwave damping.
\newblock {\em Transp. Res. C}, 148:104038, 2023.
\newblock \url{https://doi.org/10.1016/j.trc.2023.104038}.

\bibitem{Cicic2022TITS}
Mladen {\v{C}}i{\v{c}}i{\'c}, Xi~Xiong, Li~Jin, and Karl~Henrik Johansson.
\newblock Coordinating vehicle platoons for highway bottleneck decongestion and
  throughput improvement.
\newblock {\em IEEE Trans. Intell. Transp. Syst.}, 23(7):8959--8971, 2022.
\newblock \url{https://doi.org/10.1109/TITS.2021.3088775}.

\bibitem{Vishnoi2024}
Suyash~C Vishnoi, Junyi Ji, MirSaleh Bahavarnia, Yuhang Zhang, Ahmad~F Taha,
  Christian~G Claudel, and Daniel~B Work.
\newblock {CAV} traffic control to mitigate the impact of congestion from
  bottlenecks: A linear quadratic regulator approach and microsimulation study.
\newblock {\em J. Auton. Transp. Syst.}, 1(2):1--37, 2024.
\newblock \url{https://doi.org/10.1145/3636464}.

\bibitem{Goni-Ros2016}
Bernat Go{\~{n}}i-Ros, Victor~L. Knoop, Toshimichi Takahashi, Ichiro Sakata,
  Bart van Arem, and Serge~P. Hoogendoorn.
\newblock {Optimization of traffic flow at freeway sags by controlling the
  acceleration of vehicles equipped with in-car systems}.
\newblock {\em Transp. Res. C}, 71:1--18, 2016.
\newblock \url{https://doi.org/10.1016/j.trc.2016.06.022}.

\bibitem{Wang2016TITS}
Meng Wang, Winnie Daamen, Serge~P Hoogendoorn, and Bart {van Arem}.
\newblock Cooperative car-following control: Distributed algorithm and impact
  on moving jam features.
\newblock {\em IEEE Trans. Intell. Transp. Syst.}, 17(5):1459--1471, 2016.
\newblock \url{https://doi.org/10.1109/TITS.2015.2505674}.

\bibitem{Lee2025}
Jonathan~W. Lee, Han Wang, Kathy Jang, Amaury Hayat, Matthew Bunting, Arwa
  Alanqary, William Barbour, Zhe Fu, Xiaoqian Gong, George Gunter, Sharon
  Hornstein, Abdul~Rahman Kreidieh, Nathan Lichtl{\'e}, Matthew~W. Nice,
  William~A. Richardson, Adit Shah, Eugene Vinitsky, Fangyu Wu, Shengquan
  Xiang, Sulaiman Almatrudi, Fahd Althukair, Rahul Bhadani, Joy Carpio, Raphael
  Chekroun, Eric Cheng, Maria~Teresa Chiri, Fang-Chieh Chou, Ryan Delorenzo,
  Marsalis Gibson, Derek Gloudemans, Anish Gollakota, Junyi Ji, Alexander
  Keimer, Nour Khoudari, Malaika Mahmood, Mikail Mahmood, Hossein Nick~Zinat
  Matin, Sean Mcquade, Rabie Ramadan, Daniel Urieli, Xia Wang, Yanbing Wang,
  Rita Xu, Mengsha Yao, Yiling You, Gergely Zach{\'a}r, Yibo Zhao, Mostafa
  Ameli, Mirza~Najamuddin Baig, Sarah Bhaskaran, Kenneth Butts, Manasi Gowda,
  Caroline Janssen, John Lee, Liam Pedersen, Riley Wagner, Zimo Zhang, Chang
  Zhou, Daniel~B. Work, Benjamin Seibold, Jonathan Sprinkle, Benedetto Piccoli,
  Maria Laura~Delle Monache, and Alexandre~M. Bayen.
\newblock Traffic control via connected and automated vehicles ({CAVs}): An
  open-road field experiment with 100 {CAVs}.
\newblock {\em IEEE Control Syst.}, 45(1):28--60, 2025.
\newblock \url{https://doi.org/10.1109/MCS.2024.3498552}.

\bibitem{Jang2025}
Kathy Jang, Nathan Lichtl{\'e}, Eugene Vinitsky, Adit Shah, Matthew Bunting,
  Matthew Nice, Benedetto Piccoli, Benjamin Seibold, Daniel~B Work, Maria~Laura
  Delle~Monache, Jonathan Sprinkle, Jonathan~W Lee, and Alexandre~M Bayen.
\newblock Reinforcement learning-based oscillation dampening: Scaling up
  single-agent reinforcement learning algorithms to a 100-autonomous-vehicle
  highway field operational test.
\newblock {\em IEEE Control Syst.}, 45(1):61--94, 2025.
\newblock \url{https://doi.org/10.1109/MCS.2024.3503372}.

\bibitem{Treiber2013}
Martin Treiber and Arne Kesting.
\newblock {\em Traffic flow dynamics}.
\newblock Springer, 2013.
\newblock \url{https://doi.org/10.1007/978-3-642-32460-4}.

\bibitem{Treiber2000}
Martin Treiber, Ansgar Hennecke, and Dirk Helbing.
\newblock Congested traffic states in empirical observations and microscopic
  simulations.
\newblock {\em Phys. Rev. E}, 62(2):1805--1824, 2000.
\newblock \url{https://doi.org/10.1103/PhysRevE.62.1805}.

\bibitem{Treiber2015}
Martin Treiber and Venkatesan Kanagaraj.
\newblock Comparing numerical integration schemes for time-continuous
  car-following models.
\newblock {\em Phys. A}, 419:183--195, 2015.
\newblock \url{https://doi.org/10.1016/j.physa.2014.09.061}.

\bibitem{Ge2014}
Jin~I. Ge and G{\'a}bor Orosz.
\newblock Dynamics of connected vehicle systems with delayed acceleration
  feedback.
\newblock {\em Transp. Res. C}, 46:46--64, 2014.
\newblock \url{https://doi.org/10.1016/j.trc.2014.04.014}.

\bibitem{Monteil2019}
Julien Monteil, M{\'e}lanie Bouroche, and Douglas~J Leith.
\newblock $\mathcal{L}_2$ and $\mathcal{L}_{\infty}$ stability analysis of
  heterogeneous traffic with application to parameter optimization for the
  control of automated vehicles.
\newblock {\em IEEE Trans. Control Syst. Technol.}, 27(3):934--949, 2019.
\newblock \url{https://doi.org/10.1109/TCST.2018.2808909}.

\bibitem{Montanino2021a}
Marcello Montanino and Vincenzo Punzo.
\newblock On string stability of a mixed and heterogeneous traffic flow: A
  unifying modelling framework.
\newblock {\em Transp. Res. B}, 144:133--154, 2021.
\newblock \url{https://doi.org/10.1016/j.trb.2020.11.009}.

\bibitem{Wilson2008}
R~Eddie Wilson.
\newblock Mechanisms for spatio-temporal pattern formation in highway traffic
  models.
\newblock {\em Philos. Trans. R. Soc. A}, 366(1872):2017--2032, 2008.
\newblock \url{https://doi.org/10.1098/rsta.2008.0018}.

\bibitem{Montanino2021b}
Marcello Montanino, Julien Monteil, and Vincenzo Punzo.
\newblock From homogeneous to heterogeneous traffic flows: $\mathscr{L}_p$
  string stability under uncertain model parameters.
\newblock {\em Transp. Res. B}, 146:136--154, 2021.
\newblock \url{https://doi.org/10.1016/j.trb.2021.01.009}.

\bibitem{Cappiello2002}
Alessandra Cappiello, Ismail Chabini, Edward~K. Nam, Alessandro Lu{\`{e}}, and
  Maya {Abou Zeid}.
\newblock {A statistical model of vehicle emissions and fuel consumption}.
\newblock In {\em Proc. IEEE 5th Int. Conf. Intell. Transp. Syst.}, pages
  801--809. IEEE, 2002.
\newblock \url{https://doi.org/10.1109/ITSC.2002.1041322}.

\bibitem{Barth2000}
Matthew Barth, Feng An, Theodore Younglove, George Scora, Carrie Levine, Marc
  Ross, and Thomas Wenzel.
\newblock Development of a comprehensive modal emissions model, 2000.
\newblock {NCHRP} Web-Only Document 122, Contractor's Final Report for NCHRP
  Project 25-11, Transportation Research Board,
  \url{https://onlinepubs.trb.org/onlinepubs/nchrp/nchrp_w122.pdf}.

\bibitem{Hayward1972}
John~C Hayward.
\newblock Near-miss determination through use of a scale of danger.
\newblock {\em Highway Res. Rec.}, 384:24--34, 1972.

\bibitem{Kiefer2005}
Raymond~J Kiefer, David~J LeBlanc, and Carol~A Flannagan.
\newblock Developing an inverse time-to-collision crash alert timing approach
  based on drivers' last-second braking and steering judgments.
\newblock {\em Accid. Anal. Prev.}, 37(2):295--303, 2005.
\newblock \url{https://doi.org/10.1016/j.aap.2004.09.003}.

\bibitem{Schakel2010}
Wouter~J Schakel, Bart {van Arem}, and Bart~D Netten.
\newblock Effects of cooperative adaptive cruise control on traffic flow
  stability.
\newblock In {\em 13th Int. IEEE Conf. Intell. Transp. Syst.}, pages 759--764.
  IEEE, 2010.
\newblock \url{https://doi.org/10.1109/ITSC.2010.5625133}.

\bibitem{Shen2025}
Jin Shen, Jiandong Zhao, Zhixin Yu, Shiteng Zheng, and Rui Jiang.
\newblock The elimination and absorption mechanism of oscillatory motion wave
  based on jam-absorption driving for mixed traffic flow in intelligent
  connected environment.
\newblock {\em Phys. A}, 664:130485, 2025.
\newblock \url{https://doi.org/10.1016/j.physa.2025.130485}.

\bibitem{Monteil2016}
Julien Monteil and M{\'e}lanie Bouroche.
\newblock Robust parameter estimation of car-following models considering
  practical non-identifiability.
\newblock In {\em 2016 IEEE 19th Int. Conf. Intell. Transp. Syst. (ITSC)},
  pages 581--588. IEEE, 2016.
\newblock \url{https://doi.org/10.1109/ITSC.2016.7795612}.

\bibitem{Ge2018}
Jin~I. Ge and G{\'a}bor Orosz.
\newblock Connected cruise control among human-driven vehicles:
  Experiment-based parameter estimation and optimal control design.
\newblock {\em Transp. Res. C}, 95:445--459, 2018.
\newblock \url{https://doi.org/10.1016/j.trc.2018.07.021}.

\bibitem{Li2020}
Li~Li, Rui Jiang, Zhengbing He, Xiqun~Michael Chen, and Xuesong Zhou.
\newblock Trajectory data-based traffic flow studies: A revisit.
\newblock {\em Transp. Res. C}, 114:225--240, 2020.
\newblock \url{https://doi.org/10.1016/j.trc.2020.02.016}.

\bibitem{Wang2022}
Yanbing Wang, Maria~Laura Delle~Monache, and Daniel~B Work.
\newblock Identifiability of car-following dynamics.
\newblock {\em Phys. D}, 430:133090, 2022.
\newblock \url{https://doi.org/10.1016/j.physd.2021.133090}.

\bibitem{Krajewski2018}
Robert Krajewski, Julian Bock, Laurent Kloeker, and Lutz Eckstein.
\newblock The {highD} dataset: A drone dataset of naturalistic vehicle
  trajectories on {German} highways for validation of highly automated driving
  systems.
\newblock In {\em 2018 21st Int. Conf. Intell. Transp. Syst. (ITSC)}, pages
  2118--2125. IEEE, 2018.
\newblock \url{https://doi.org/10.1109/ITSC.2018.8569552}.

\bibitem{Seo2021}
Toru Seo, Yusuke Tago, Norihito Shinkai, Masakazu Nakanishi, Jun Tanabe,
  Daisuke Ushirogochi, Shota Kanamori, Atsushi Abe, Takashi Kodama, Satoshi
  Yoshimura, Masaaki Ishihara, and Wataru Nakanishi.
\newblock Evaluation of large-scale complete vehicle trajectories dataset on
  two kilometers highway segment for one hour duration: {Zen Traffic Data}.
\newblock In {\em 2021 Int. Symp. Transp. Data Model. (ISTDM2021)}. University
  of Michigan, 2021.
\newblock
  \url{https://limos.engin.umich.edu/istdm2021/wp-content/uploads/sites/2/2021/05/ISTDM-2021-Extended-Abstract-0019.pdf}.

\bibitem{Gloudemans2023}
Derek Gloudemans, Yanbing Wang, Junyi Ji, Gergely Zachar, William Barbour, Eric
  Hall, Meredith Cebelak, Lee Smith, and Daniel~B Work.
\newblock I-24 motion: An instrument for freeway traffic science.
\newblock {\em Transp. Res. C}, 155:104311, 2023.
\newblock \url{https://doi.org/10.1016/j.trc.2023.104311}.

\end{thebibliography}


\end{document}